\documentclass[journal,10pt]{IEEEtran}
\usepackage[T1]{fontenc}
\usepackage{microtype}
\usepackage[utf8]{inputenc}
\usepackage{amsmath,amsfonts,amssymb}
\usepackage{array,booktabs,multirow,cite,textcomp}
\usepackage[caption=false,font=footnotesize,labelfont=sf,textfont=sf,labelformat=simple]{subfig}
\usepackage{stfloats,url,graphicx}
\allowdisplaybreaks
\makeatletter
\def\ftype@table{1}
\def\@IEEEtablecaptionsepspace{\vskip2pt\relax}
\renewcommand{\section}{\@startsection{section}{1}{\z@}
  {2ex}{0.7ex}{\normalfont\normalsize\centering\scshape}}
\renewcommand{\subsection}{\@startsection{subsection}{2}{\z@}
  {2ex}{0.7ex}{\normalfont\normalsize\itshape}}
\makeatother
\begin{document}
\bstctlcite{Paper2ReferenceStyle}
\flushbottom
\setlength{\parskip}{0pt plus 1pt}
\setlength{\emergencystretch}{1em}
\renewcommand{\thesubfigure}{(\alph{subfigure})}
\setlength{\abovedisplayskip}{2pt plus 2pt minus 1pt}
\setlength{\belowdisplayskip}{2pt plus 2pt minus 1pt}
\setlength{\abovedisplayshortskip}{2pt plus 1pt minus 1pt}
\setlength{\belowdisplayshortskip}{2pt plus 1pt minus 1pt}

\title{LC-Resonance Placement Limits and Damping Recovery in Virtual-Admittance-Based Grid-Forming Inverter Control}

\author{Stephen Arinze Obi,~\IEEEmembership{Graduate Student Member,~IEEE},~and Jae-Jung Jung,~\IEEEmembership{Senior Member,~IEEE}
\thanks{The authors are with the School of Electronics and Electrical Engineering,
Kyungpook National University, Daegu 41566, Republic of Korea
(e-mail: mashall96@knu.ac.kr; jj.jung@knu.ac.kr).}
\thanks{Corresponding author: Jae-Jung Jung.}
}

\maketitle

\begin{abstract}

Compact LC-filter design and inner-loop control must be coordinated for stable grid-forming operation. This paper quantifies how current-loop proportional gain constrains LC-resonance placement in virtual-admittance-based dual-loop control with unity voltage-feedback decoupling (VFD) and capacitor-voltage active damping. Equivalent-impedance analysis identifies gain-dependent contraction of the positive-resistance region. Delayed inverter-current feedback opposes VFD-assisted damping above one-sixth of the sampling frequency. Simple proportional-path frequency shaping recovers an admissible effective-gain window while retaining unity VFD and leaving the resonant path unchanged. Reduced-model criteria, checked against the complete model, link window closure to a nominal resonance-placement limit near one-third of the sampling frequency under single-sampled single-update operation over the studied gain range. A 3-kVA prototype demonstrates an 83.3\% LC-product reduction at comparable voltage and nominal current crossovers. A separate comparison with the same high-resonance filter demonstrates stable operation at 2.68 times the unshaped nominal current crossover. Output-impedance analysis and grid-connected tests identify additional grid-dependent gain constraints.

\end{abstract}

\begin{IEEEkeywords}

Active damping, frequency shaping, grid-forming inverters, LC resonance, virtual admittance, voltage-feedback decoupling.

\end{IEEEkeywords}

\section{Introduction}

The increasing integration of renewable generation and energy storage is expanding the role of grid-forming inverters (GFMIs), whose controlled voltage-source behavior supports voltage and frequency in converter-rich networks \cite{Rosso2021GridFormingReview}. Their stability depends on grid interaction \cite{Zhang2025EnergyStorageInteraction}, \cite{Qu2026AdaptiveStabilityBoundaries} and the inner loops that realize the commanded voltage-source behavior \cite{Zhao2024ClosedFormDesign}. Inadequate output-filter damping can compromise voltage regulation and grid-connected operation. Meanwhile, compact power-conversion systems favor smaller passive components, making coordinated filter selection and controller tuning essential for sustainable grid operation.

Dual-loop control uses inner current control (CC) and outer voltage control (VC) to realize the power-control layer (PCL) voltage command \cite{Zhao2021CurrentLoopTuning}, \cite{He2023DissipativityRobustness}. In virtual-admittance (VA)-based control, a virtual resistance--inductance network converts voltage error into a current reference \cite{Liao2020PassivityVoltageControllers,Zhao2025InnerLoopDamping,He2024HighFrequencySTATCOM}. The inverter-side current loop provides inherent damping while regulating filter-inductor current, the sum of capacitor current and current delivered to the load or grid \cite{Geng2018InductorCurrentDamping}, \cite{Tang2012InherentDamping}. In contrast, grid-side current control regulates delivered current and has distinct damping characteristics, including delay-induced damping in LCL configurations \cite{Wang2016GridCurrentDamping}. This paper considers inverter-side current control.

Capacitor-voltage feedforward, termed voltage-feedback decoupling (VFD), compensates capacitor-voltage coupling in the inductor-current dynamics. Feeding the capacitor voltage directly to the current-controller output improves startup transients and disturbance rejection \cite{Liao2020PassivityVoltageControllers}, \cite{DeBosio2016StateFeedbackCoupling}. Under inductive VA operation, however, unity VFD cancels the voltage loop's low-frequency restoring term, making stability sensitive to digital control delay. Attenuating VFD restores this term but couples voltage crossover to current gain. Co-designing a stability-constrained VFD gain and virtual impedance improves internal stability and terminal passivity \cite{ObiVirtualAdmittanceVFD}. The architecture considered here adopts negative proportional capacitor-voltage feedback in the current reference to provide active damping (AD), restoring voltage-loop support while retaining unity VFD \cite{Obi2026VFDTrapPreprint}. With sufficient bandwidth separation and dominant AD support, its gain approximately sets voltage crossover independently of current gain.

The resulting low-frequency regulation must remain compatible with LC-resonance damping. Digital control delay couples current gain to filter selection through the damping contributions of inverter-current feedback and VFD \cite{Wang2014DigitalLCLDesign,Xia2017ControllerDelays,RodriguezDiaz2019InputAdmittancePassivity}. Under single-sampled single-update pulsewidth modulation (PWM), a combined computation and PWM delay of 1.5 sampling periods causes proportional inverter-current feedback to lose positive equivalent resistance at $f_s/6$, where $f_s$ is the sampling frequency \cite{Geng2018InductorCurrentDamping}. The positive virtual-resistance region (PVRR) is the interval of positive equivalent resistance. In single-loop LC voltage control, positive proportional capacitor-voltage feedback extends it toward $f_s/3$ \cite{Zhang2022CapacitorVoltageDamping}. Feedforward-damping studies also relate LC-filter constraints and passivity to PWM delay and sampling arrangements \cite{He2025FeedforwardDamping}. Extending the region supports compact filters: higher physical resonance permits a smaller inductance--capacitance product, subject to harmonic attenuation and component-stress requirements.

Capacitor-current, capacitor-voltage, inverter-current, and grid-current feedback provide AD in LC- and LCL-filtered converters \cite{Dannehl2010ActiveDampingApproaches,Dannehl2011FilterBasedDamping,Parker2012ActiveDampingRegions,Wang2015VirtualRCDamping}. In LCL converters with grid-side current control, phase compensation and capacitor-current differential feedback extend the PVRR and improve robustness to grid-inductance variation \cite{Fang2021PhaseLeadDamping,Ding2022DelayCompensation,Ke2023RobustActiveDamping,Woo2026DifferentialCapacitorCurrent}. Furthermore, low-pass and all-pass filters shape the damping response, while generalized design methods relate the achievable region to feedback gain and filter phase \cite{Cai2018LowPassSingleLoop,RoldanPerez2018AllPassDamping,Yao2020PhaseReshaping,Wang2023GraphicalDampingDesign}. Low-pass-based current control coordinated with voltage feedforward also improves terminal passivity \cite{Chen2020DominantAdmittance}. In VA-based grid-forming control, impedance shaping and admittance synthesis address inner-loop damping and grid interaction \cite{Zhao2025InnerLoopDamping}, \cite{He2024HighFrequencySTATCOM}, \cite{Li2026VirtualAdmittanceVSG}, \cite{Cvetanovic2026AdmittanceSynthesis}. The AD-supported unity-VFD architecture and low-frequency tuning are established in \cite{Obi2026VFDTrapPreprint}; the remaining question is how current gain and capacitor-voltage AD constrain LC-resonance placement and how to recover an admissible gain window.

This paper shows that increasing current-loop proportional gain strengthens adverse delayed inverter-current feedback above $f_s/6$, contracts the combined PVRR, and can destabilize a resonance supported at lower gain. Reducing the gain recovers damping but lowers current crossover and loop separation. We therefore adopt simple proportional-path frequency shaping using a first-order low-pass filter (LPF), retaining unity VFD and leaving the resonant path unchanged. Coordinated filter and inner-loop design restores design freedom under tight damping and bandwidth constraints. The main contributions are as follows.

\begin{enumerate}

\item An equivalent-impedance analysis establishes how increasing current-loop proportional gain contracts the PVRR in the considered unity-VFD, VA-based dual-loop architecture and explains damping recovery through simple proportional-path frequency shaping.
\item Delay-aware reduced-model criteria quantify the two-sided effective-gain window and link its closure to the physical LC-resonance placement limit. Predictions are checked against the complete inner-loop analytical model, whose poles quantify modal damping.
\item Experiments on a 3-kVA prototype demonstrate design freedom: a smaller LC product at comparable crossover targets or higher current gain with the same filter. Unloaded tests corroborate both AD-gain limits; grid-connected tests support impedance-based constraints at selected weak- and strong-grid operating points.

\end{enumerate}

\section{Control Structure and LC-Resonance Damping Mechanism}

\subsection{System and Inner Control Structure}

Fig. 1 shows the three-phase LC-filtered VA-based GFMI. The power-synchronization controller (PSC) sets the internal voltage source (IVS) angle $\theta_c$, while the reactive-power controller (RPC) adjusts its magnitude $E$, giving $e^*=E\angle\theta_c$. The PSC generates $\Delta\omega_o=D_{\mathrm p}(P^*-P)$, while the RPC gives $\Delta E=D_{\mathrm q}(Q^*-Q)/s$, where $P$ and $Q$ are the low-pass-filtered power measurements. Their lower bandwidths allow PCL dynamics to be neglected throughout the small-signal analysis \cite{Qu2026AdaptiveStabilityBoundaries}, \cite{Liao2020PassivityVoltageControllers}. The symmetric stationary-frame inner control is therefore represented by a single-input single-output (SISO) model. Including PCL dynamics generally introduces coupling requiring a multiple-input multiple-output (MIMO) model.

\begin{figure}[!t]
\centering
\includegraphics[width=3.318in]{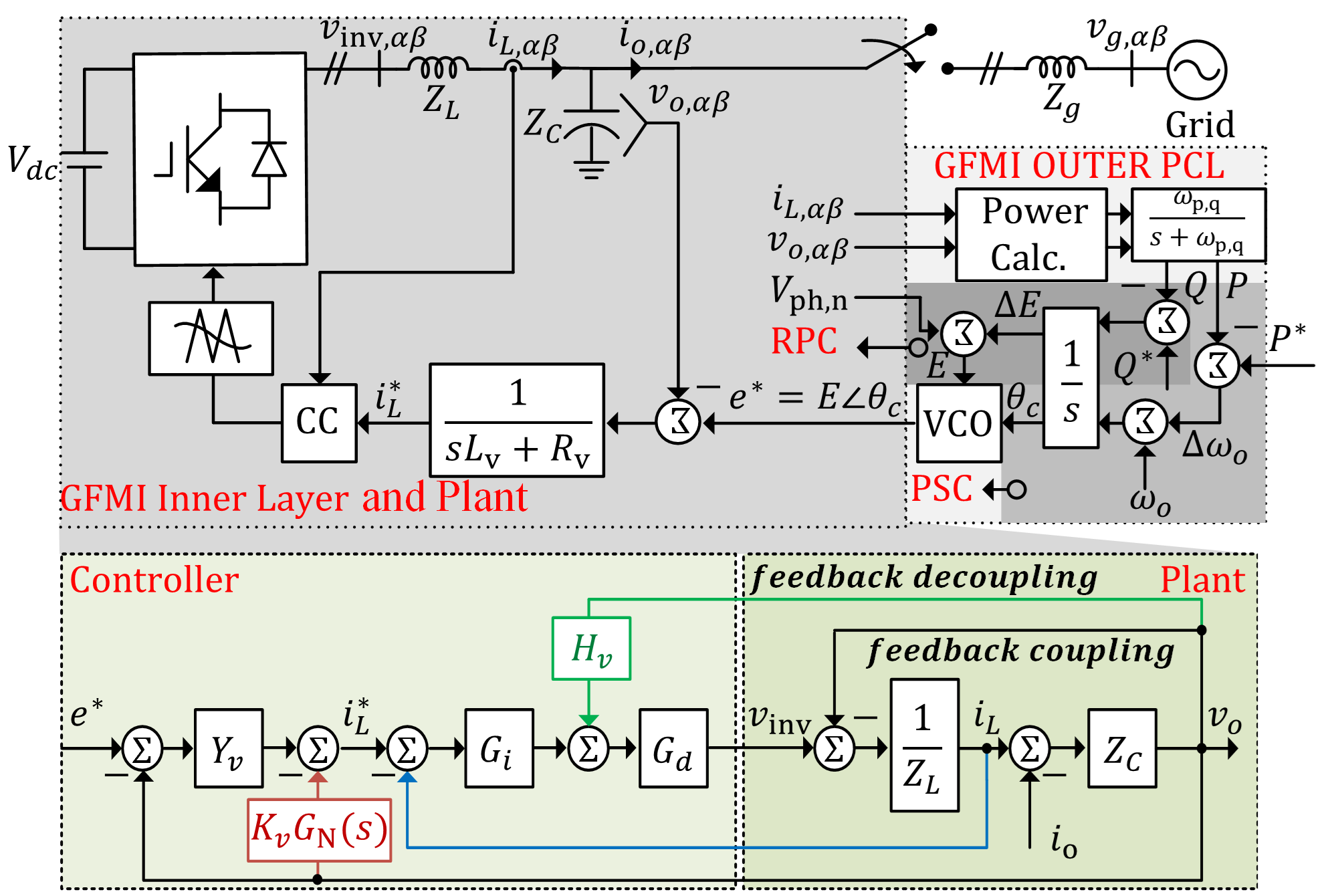}
\par\vspace{-5.5pt}
\caption{LC-filtered VA-GFMI with unity VFD, capacitor-voltage AD, and proportional-path shaping. VCO: voltage-controlled oscillator.}
\label{fig:1}
\end{figure}

For one stationary-frame control axis, ${v}_{o}$ denotes the capacitor voltage and ${i}_{L}$ the inverter-side filter-inductor current. The physical filter impedances are ${Z}_{L}(s)=s{L}_{f}$ and ${Z}_{C}(s)=1/(s{C}_{f})$, giving the physical resonance ${f}_{r}=1/(2\pi \sqrt{{L}_{f}{C}_{f}})$. The analytical models neglect inductor winding resistance, capacitor equivalent series resistance, and cable resistance. The virtual admittance converts the voltage error into a current reference through

\begin{equation}{Y}_{\mathrm{v}}(s)=1/({R}_{\mathrm{v}}+s{L}_{\mathrm{v}}),\tag{1}\label{eq:1}\end{equation}
where ${R}_{\mathrm{v}}$ and ${L}_{\mathrm{v}}$ are the virtual resistance and inductance. Proportional capacitor-voltage AD is included in the current reference as

\begin{equation}\begin{aligned}
{i}_{L}^{*}&={Y}_{\mathrm{v}}\left( s\right) \left( {e}^{*}-{v}_{o}\right) -{K}_{v}{G}_{\mathrm{N}}\left( s\right) {v}_{o},\\
{G}_{\mathrm{N}}(s)&=\frac{{s}^{2}+{\omega }_{o}^{2}}{{s}^{2}+2{\zeta }_{\mathrm{N}}{\omega }_{o}s+{\omega }_{o}^{2}}.
\end{aligned}\tag{2}\label{eq:2}\end{equation}
Here, $K_v>0$ is the AD gain and $\omega_o=2\pi f_o$ is the fundamental angular frequency. The AD feedback restores the voltage-loop support cancelled by unity VFD and enables voltage-bandwidth tuning under sufficient bandwidth separation. The notch filter $G_{\mathrm N}(s)$, with $\zeta_{\mathrm N}=0.1$, suppresses the fundamental component in this path.

The proportional-resonant (PR) current controller is

\begin{equation}\begin{aligned}
{G}_{i}\left( s\right) &={K}_{pi}{G}_{\mathrm{LPF}}\left( s\right) +\frac{2{K}_{ri}{\omega }_{i}s}{{s}^{2}+2{\omega }_{i}s+{\omega }_{o}^{2}},\\
{G}_{\mathrm{LPF}}(s)&={\omega}_{\mathrm{LPF}}/(s+{\omega}_{\mathrm{LPF}}).
\end{aligned}\tag{3}\label{eq:3}\end{equation}
Here, $K_{pi}$ and $K_{ri}$ are proportional and resonant gains, and $\omega_i$ is the resonant-bandwidth parameter. Only the proportional path is shaped, with $\omega_{\mathrm{LPF}}=2\pi f_{\mathrm{LPF}}$; the resonant branch provides fundamental tracking. At $K_{ri,\mathrm{p.u.}}=5$, the proportional path dominates the examined nominal-crossover and LC-resonance ranges, supporting the approximation below.

The commanded and applied inverter voltages are

\begin{equation}\begin{aligned}
u_c^*&={G}_{i}(s)\left[{i}_{L}^{*}-{i}_{L}\right]+{H}_{v}{v}_{o},\\
{v}_{\mathrm{inv}}&={G}_{d}(s)u_c^*.
\end{aligned}\tag{4}\label{eq:4}\end{equation}
Feedforward enters at the current-controller output, whereas AD enters through the current reference; consequently, the two paths experience different controller dynamics. The combined digital control delay is $G_d(s)=e^{-sT_d}$. For the adopted single-sampled single-update (SSSU) PWM operation, $f_s=f_{sw}$, $T_s=1/f_s$, and $T_d=1.5T_s$.

Base VA--CC uses $H_v=K_v=0$ and $G_{\mathrm{LPF}}=1$; unshaped VFD--AD retains unity VFD and AD with $G_{\mathrm{LPF}}=1$. The proposed frequency-shaped controller adds proportional-path shaping and is labeled ``Proposed'' in the experimental results.

The impedance base is $Z_{\mathrm b}=V_{\mathrm{LL,b}}^2/S_{\mathrm b}$, using the rated line-to-line root-mean-square (RMS) voltage and three-phase apparent power in Table I. Phase waveforms use peak bases $V_{\mathrm b}=\sqrt{2/3}V_{\mathrm{LL,b}}$ and $I_{\mathrm b}=\sqrt{2/3}S_{\mathrm b}/V_{\mathrm{LL,b}}$. The per-unit (p.u.) gains are $K_{pi,\mathrm{p.u.}}=K_{pi}/Z_{\mathrm b}$, $K_{ri,\mathrm{p.u.}}=K_{ri}/Z_{\mathrm b}$, and $K_{v,\mathrm{p.u.}}=K_vZ_{\mathrm b}$. The dimensionless effective AD gain is $K_{\mathrm{eff}}=K_{pi}K_v=K_{pi,\mathrm{p.u.}}K_{v,\mathrm{p.u.}}$. $K_{pi}$ governs current-loop dynamics, whereas $K_v$ adjusts the voltage-loop restoring term.

\begin{table}[!t]

\caption{System and Control Parameters}
\label{tab:1}
\centering\footnotesize
\setlength{\tabcolsep}{2.8pt}
\renewcommand{\arraystretch}{1.0}
\begin{tabular}{@{}l p{.36\columnwidth} p{.25\columnwidth} p{.13\columnwidth}@{}}
\toprule
Symbol & Description & Value (SI) & p.u. \\
\midrule
${S}_{\mathrm{b}}$ & Rated three-phase apparent power & $3\,\mathrm{kVA}$ & 1.0 \\
${V}_{\mathrm{LL,b}}$ & Rated line-to-line RMS & $110\,\mathrm{V}$ & 1.0 \\
${V}_{\mathrm{dc}}$ & DC-link voltage & $300\,\mathrm{V}$ & --- \\
${f}_{o}$ & Fundamental frequency & 50~Hz & 1.0 \\
${Z}_{\mathrm{b}}$ & Base impedance & 4.033 \ensuremath{\Omega} & 1.0 \\
${f}_{sw},{f}_{s}$ & Switching/sampling freq. & $10,10\,\mathrm{kHz}$ & --- \\
${\omega }_{\mathrm p},{\omega }_{\mathrm q}$ & PSC/RPC power-filter freq. & $10\pi\,\mathrm{rad/s}$ & --- \\
${D}_{\mathrm{p}}$ & PSC gain & \shortstack{$2.1\times10^{-3}$\\$\mathrm{rad}/(\mathrm{W}\,\mathrm{s})$} & --- \\
${D}_{\mathrm{q}}$ & RPC integral gain & \shortstack{$70\times10^{-3}$\\$\mathrm{V}/(\mathrm{var}\,\mathrm{s})$} & --- \\
${R}_{\mathrm{v}}$ & Virtual resistance & $0.807\,\Omega$ & 0.200 \\
${L}_{\mathrm{v}}$ & Virtual inductance & $4.49\,\mathrm{mH}$ & 0.350 \\
${K}_{ri}$ & Resonant CC gain & $20.2\,\Omega$ & 5.0 \\
${\omega }_{i}$ & Resonant bandwidth & $2\pi\cdot1\,\mathrm{rad/s}$ & --- \\
${H}_{v}$ & VFD gain & 1 & 1.0 \\
\bottomrule
\end{tabular}
\end{table}

\subsection{Capacitor-Node Equivalent Impedance}

Fig. 2 represents the feedback paths as parallel equivalent impedances at the capacitor node. Under no-load conditions, $i_o=0$, $i_L=sC_fv_o$. Using the resonance-band approximations $G_i(s)\approx K_{pi}G_{\mathrm{LPF}}(s)$ and $G_{\mathrm N}(s)\approx1$, the current-feedback, VFD, and capacitor-voltage AD impedances are given by (5).

\begin{figure}[!t]
\centering
\subfloat[]{\includegraphics[width=3.044in]{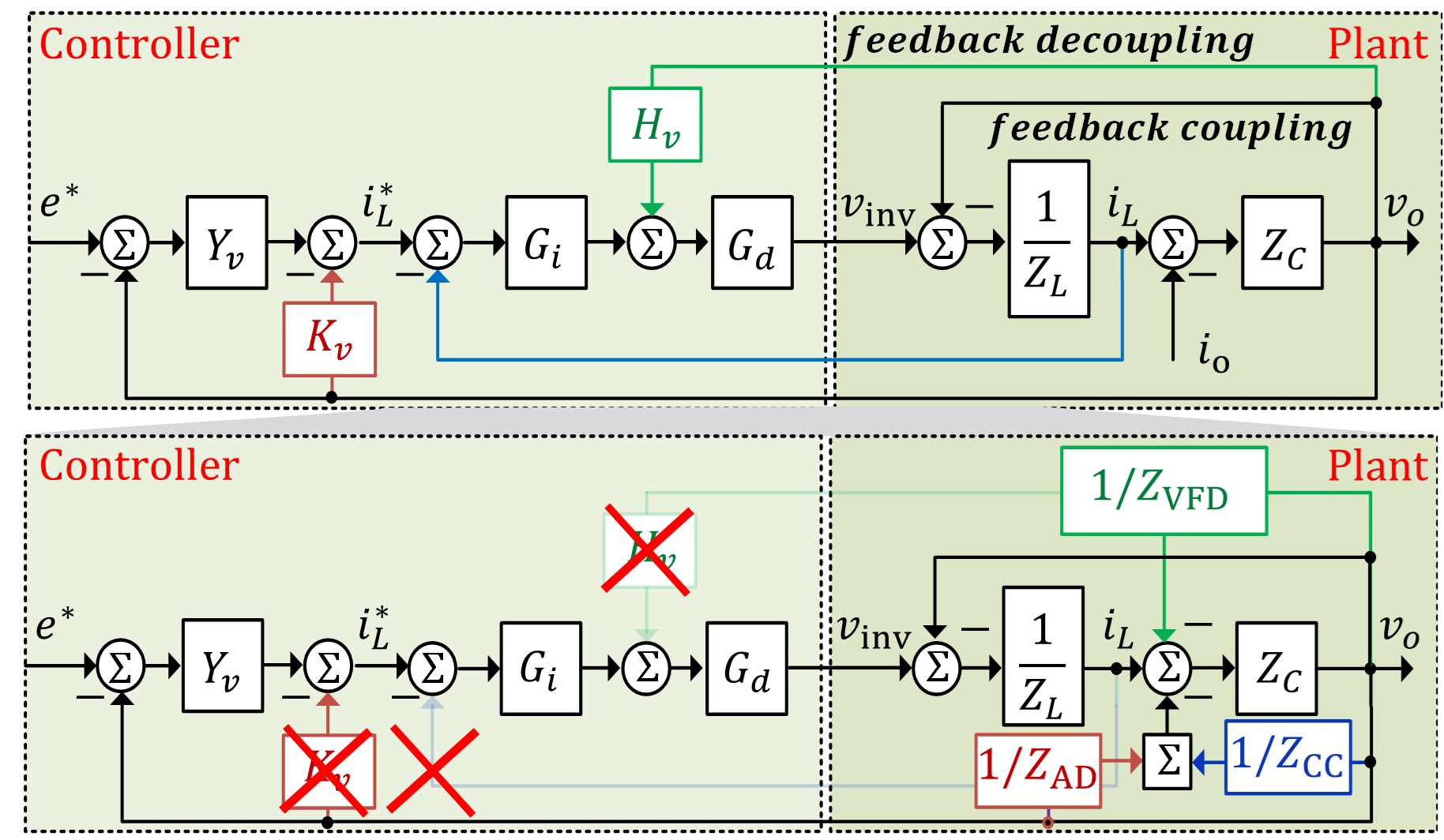}}
\par\vspace{1pt}
\subfloat[]{\includegraphics[width=2.416in]{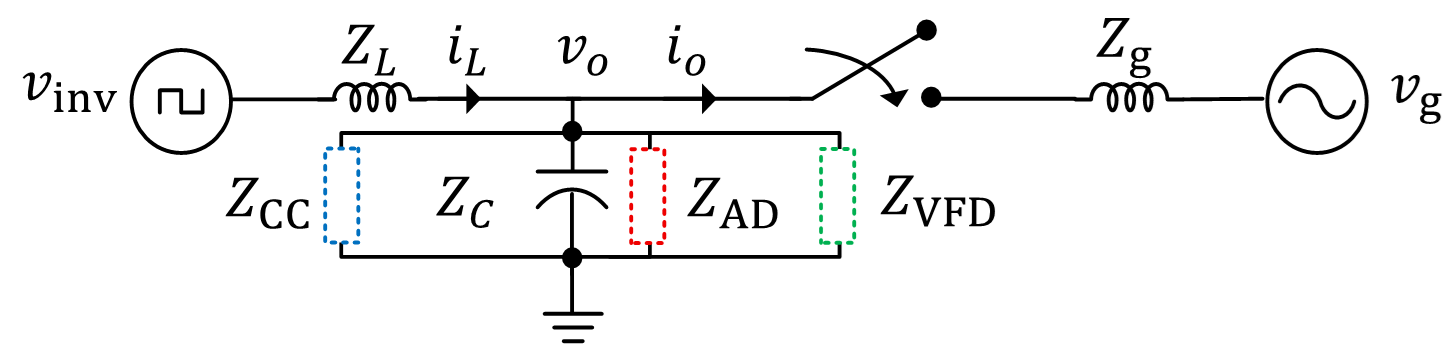}}
\caption{Resonance-band representation: (a) feedback paths at the capacitor node and (b) equivalent circuit.}
\label{fig:2}
\end{figure}

\begin{equation}\begin{aligned}
{Z}_{\mathrm{CC}}(s)&=\frac{Z_LZ_C}{G_dK_{pi}G_{\mathrm{LPF}}},\quad
Z_{\mathrm{VFD}}(s)=-\frac{Z_L}{G_dH_v},\\
{Z}_{\mathrm{AD}}(s)&=\frac{Z_L}{G_dK_{\mathrm{eff}}G_{\mathrm{LPF}}}.
\end{aligned}\tag{5}\label{eq:5}\end{equation}
Their parallel combination gives

\begin{equation}{Z}_{\mathrm{eq}}(s)={Z}_{\mathrm{CC}}(s)\parallel {Z}_{\mathrm{VFD}}(s)\parallel {Z}_{\mathrm{AD}}(s).\tag{6}\label{eq:6}\end{equation}
Writing $Z_{\mathrm{eq}}(j\omega)=R_{\mathrm{eq}}(\omega)+jX_{\mathrm{eq}}(\omega)$ separates the equivalent-resistance contribution to damping from feedback-induced reactance. The PVRR satisfies $R_{\mathrm{eq}}>0$. The equivalent network isolates the current-feedback, VFD, and AD contributions; Section III retains VA feedback when establishing the closed-loop stability boundaries.

The representative filter uses ${L}_{f}=0.5\,\mathrm{mH}$ and ${C}_{f}=10\,\text{\textmu F}$, giving ${f}_{r}=2.251\,\mathrm{kHz}$. Unless varied, Figs. 3--4 use $K_{pi,\mathrm{p.u.}}=0.90$, AD-enabled control uses $K_{v,\mathrm{p.u.}}=0.70$, and proposed control uses $f_{\mathrm{LPF}}=0.9f_r$. Curves are normalized by $L_f/(K_{pi}C_f)$ at $K_{pi,\mathrm{p.u.}}=0.90$.

\subsection{Current-Gain Dependence and Proportional-Path Shaping}

\subsubsection{Competing Feedback Contributions}

Fig. 3 compares the normalized equivalent resistance and reactance of base VA--CC, unshaped VFD--AD, and the frequency-shaped controller. With $\theta=\omega T_d$ and $a_{\mathrm{LPF}}=\omega/\omega_{\mathrm{LPF}}$, the current-feedback impedance in (5) gives

\begin{figure}[!t]
\centering
\subfloat[]{\includegraphics[width=3.173in]{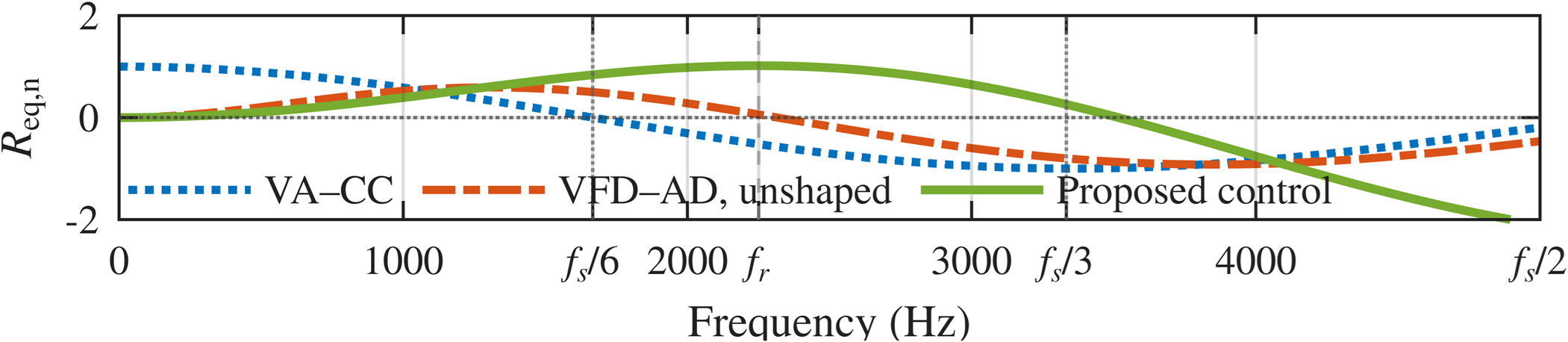}}
\par\vspace{1pt}
\subfloat[]{\includegraphics[width=3.173in]{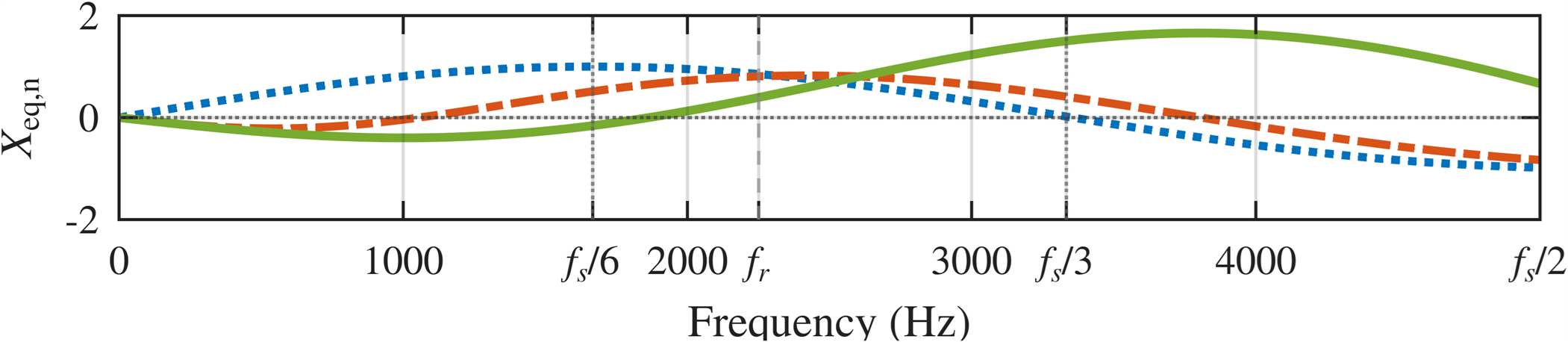}}
\caption{Normalized equivalent impedance of base VA--CC, unshaped VFD--AD, and shaped control: (a) resistance and (b) reactance.}
\label{fig:3}
\end{figure}

\begin{equation}\begin{aligned}
{R}_{\mathrm{CC}}(\omega )&=\frac{{L}_{f}}{{K}_{pi}{C}_{f}}\left( \cos(\theta) -{a}_{\mathrm{LPF}}\sin(\theta) \right) ,\\
{X}_{\mathrm{CC}}(\omega )&=\frac{{L}_{f}}{{K}_{pi}{C}_{f}}\left( {a}_{\mathrm{LPF}}\cos(\theta) +\sin(\theta) \right) .
\end{aligned}\tag{7}\label{eq:7}\end{equation}
For unshaped control, $a_{\mathrm{LPF}}=0$, and $R_{\mathrm{CC}}$ crosses zero at $\theta=\pi/2$, corresponding to $f_s/6$ for $T_d=1.5T_s$. Above this frequency, delayed inverter-current feedback contributes negative equivalent resistance. Increasing $K_{pi}$ reduces $|Z_{\mathrm{CC}}|$, strengthening its influence on the combined damping.

In contrast, the VFD branch gives

\begin{equation}\begin{aligned}
R_{\mathrm{VFD}}(\omega)=\frac{\omega L_f}{H_v}\sin\theta,\quad
X_{\mathrm{VFD}}(\omega)=-\frac{\omega L_f}{H_v}\cos\theta.
\end{aligned}\tag{8}\label{eq:8}\end{equation}
The VFD branch provides positive equivalent resistance over $0<\theta<\pi$, corresponding to $0<f<f_s/3$. Its contribution is independent of $K_{pi}$ and extends beyond the current-feedback boundary, supporting a wider combined PVRR.

The capacitor-voltage AD branch gives

\begin{equation}\begin{aligned}
{R}_{\mathrm{AD}}\left( \omega \right) &=-\frac{\omega {L}_{f}}{{K}_{pi}{K}_{v}}\left( \sin(\theta) +{a}_{\mathrm{LPF}}\cos(\theta) \right) ,\\
{X}_{\mathrm{AD}}\left( \omega \right) &=\frac{\omega {L}_{f}}{{K}_{pi}{K}_{v}}\left( \cos(\theta) -{a}_{\mathrm{LPF}}\sin(\theta) \right) .
\end{aligned}\tag{9}\label{eq:9}\end{equation}
For unshaped control, $R_{\mathrm{AD}}<0$ over $0<\theta<\pi$, so net damping follows from the parallel combination in (6). At fixed $K_v$, increasing $K_{pi}$ reduces both $|Z_{\mathrm{CC}}|$ and $|Z_{\mathrm{AD}}|$. Above $f_s/6$, these branches increasingly oppose the positive VFD contribution, contracting the combined PVRR. Proportional-path shaping increases their impedance magnitudes and modifies their phases while leaving $Z_{\mathrm{VFD}}$ unchanged.

Fig. 3(a) shows that shaping recovers positive equivalent resistance around the representative LC resonance and moves the upper PVRR boundary toward $f_s/3$, expanding resonance-placement choices. The reactance change in Fig. 3(b) also shifts the internal mode $f_r'$ relative to $f_r$. Consequently, the shaping cutoff must recover damping while maintaining internal stability and the required regulation bandwidth.

\subsubsection{LPF Cutoff and Current-Loop Dynamics}

For gain initialization, $f_{c,i}$ denotes the nominal current crossover from (11). With a slower voltage crossover, the shaping cutoff is placed above this design scale and below the physical resonance:

\begin{equation}{f}_{c,v}\approx1/(2\pi{L}_{\mathrm{v}}{K}_{v})<{f}_{c,i}<{f}_{\mathrm{LPF}}<{f}_{r}.\tag{10}\label{eq:10}\end{equation}
The voltage-crossover approximation in (10) follows the AD-supported relation derived in \cite[eq.~(17)]{Obi2026VFDTrapPreprint}, with $K_v$ denoting $K_{\mathrm{ad}}$. Its use here requires sufficient bandwidth separation and dominance of the AD contribution.

Fig. 4(a) shows that a cutoff near the physical resonance substantially extends the PVRR. Raising the cutoff retains more adverse feedback; lowering it substantially adds current-loop phase lag with little further extension. Accordingly, $f_{\mathrm{LPF}}=0.9f_r$ is selected for the following design.

\begin{figure}[!t]
\centering
\subfloat[]{\includegraphics[width=3.173in]{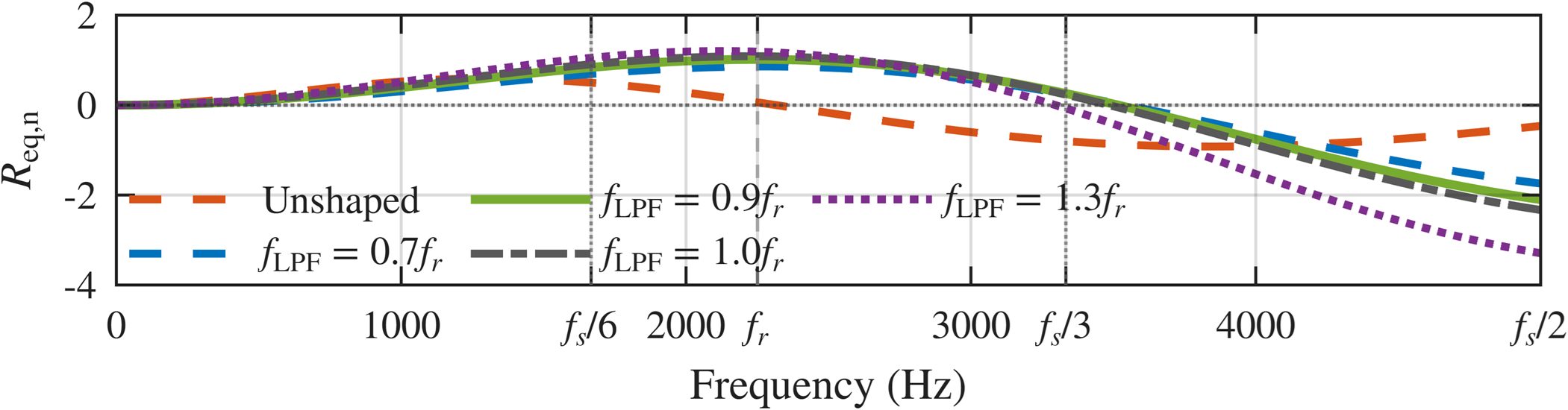}}
\par\vspace{1pt}
\subfloat[]{\includegraphics[width=3.173in]{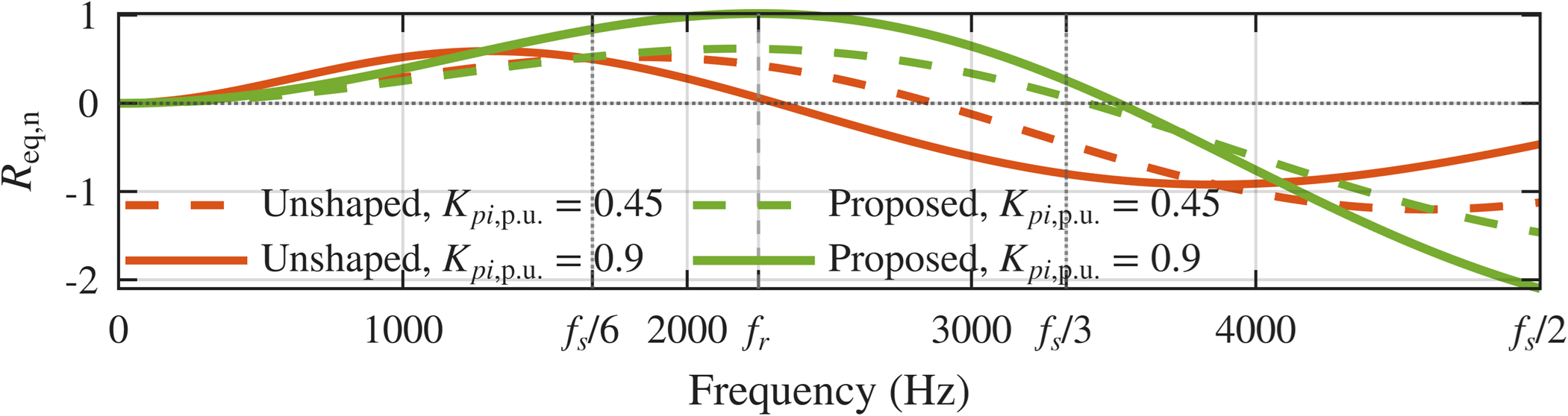}}
\caption{Effect of proportional-path shaping on the PVRR: (a) LPF-cutoff placement and (b) current-gain comparison at $K_{pi,\mathrm{p.u.}}=0.45,0.90$.}
\label{fig:4}
\end{figure}

Fig. 4(b) shows that increasing $K_{pi}$ lowers the upper PVRR boundary without shaping. Shaping retains it near $f_s/3$ over the examined range, permitting higher current gain with positive equivalent resistance around the selected resonance. The nominal inductor-loop gain quantifies the associated crossover and phase-margin tradeoff:

\begin{equation}{T}_{i}(s)={G}_{d}(s){G}_{i}(s)/(s{L}_{f}).\tag{11}\label{eq:11}\end{equation}
The unshaped approximation $f_{c,i}\approx K_{pi}/(2\pi L_f)$ initializes current gain. At fixed gain, shaping lowers this nominal crossover and adds phase lag. These quantities use the controller in (3); Appendix A gives proportional-path approximations and the current-loop expression with VFD and AD retained.

Fig. 5 compares nominal and VFD--AD current-loop gains with VA feedback opened. The latter has no right-half-plane (RHP) poles, so the Nyquist criterion requires zero net encirclements of $-1$ \cite{Liao2020ImpedanceRHPCriterion}. Shaping moves its resonance-band $-180^\circ$ crossing from $+7.91$ to $-6.40$~dB, removing two clockwise encirclements despite the lower nominal phase margin. Supplementary Fig.~S1 gives the Nyquist plots. Closing VA feedback gives the voltage-loop gain examined next.

\begin{figure}[!t]
\centering
\includegraphics[width=3.20815in]{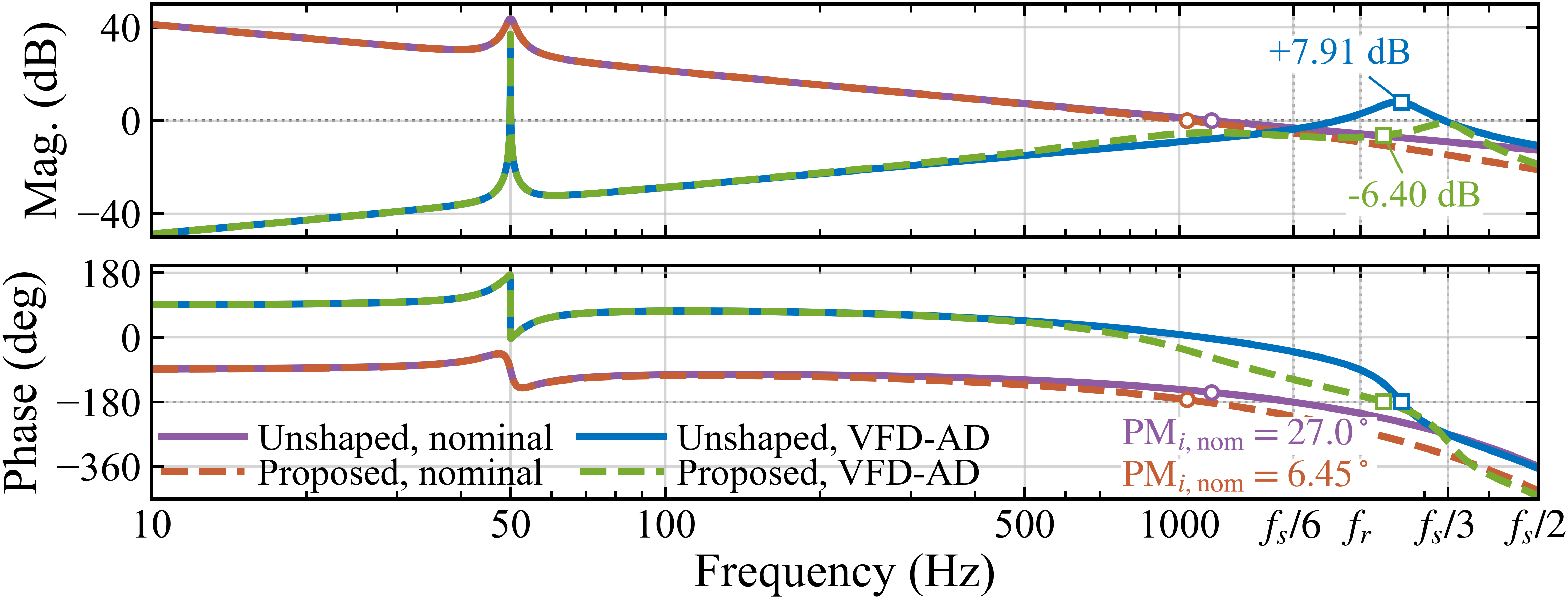}
\par\vspace{-8.6pt}
\caption{Current-loop gains for the representative filter at $K_{pi,\mathrm{p.u.}}=0.90$ and $K_{v,\mathrm{p.u.}}=0.70$. Solid/dashed: unshaped/proposed; circles/squares: nominal gain/VFD--AD phase crossings. VA feedback is opened.}
\label{fig:5}
\end{figure}

\section{Gain-Window and Physical-Resonance Design Criteria}

\subsection{Voltage-Loop Gain and Resonance-Band Reduction}

The current-loop closure in Fig. 5 supplies the voltage-loop plant. Internal stability is assessed under no-load conditions, ${i}_{L}=s{C}_{f}{v}_{o}$. Combining (1)--(4) gives the complete inner-loop analytical model used to quantify admissible gains and physical-resonance limits:

\begin{equation}\begin{aligned}
{T}_{v}(s)&=\frac{{G}_{d}(s){G}_{i}(s){Y}_{\mathrm{v}}(s)}{{D}_{v}(s)},\\
{D}_{v}(s)&=s{C}_{f}[s{L}_{f}+{G}_{d}(s){G}_{i}(s)]\\
&\quad +1-{G}_{d}(s){H}_{v}+{G}_{d}(s){G}_{i}(s){K}_{v}{G}_{\mathrm{N}}(s).
\end{aligned}\tag{12}\label{eq:12}\end{equation}
Appendix B derives (12), retaining VA resistance, the complete PR controller, AD, and digital control delay. Boundaries in Figs. 8 and 16(a) use $G_d(s)$ without approximation; RHP-root counts identify stable intervals. Third-order Pad\'{e} approximation supplies the plotted poles and modal damping.

Over the resonance and gain-boundary frequency ranges, the virtual admittance is predominantly inductive, the proportional current-control path dominates, and the AD notch response approaches unity. Thus, $Y_{\mathrm v}(s)\approx1/(sL_{\mathrm v})$, $G_i(s)\approx K_{pi}G_{\mathrm{LPF}}(s)$, and $G_{\mathrm N}(s)\approx1$. With unity VFD and $K_{\mathrm{eff}}=K_{pi}K_v$, the reduced voltage-loop gain is

\begin{equation}\begin{aligned}
{T}_{v}^{'}(s)&=\frac{{G}_{d}(s){K}_{pi}{G}_{\mathrm{LPF}}(s)}{s{L}_{\mathrm{v}}{D}_{v}^{'}(s)},\\
{D}_{v}^{'}(s)&=s{C}_{f}[s{L}_{f}+{G}_{d}(s){K}_{pi}{G}_{\mathrm{LPF}}(s)]\\
&\quad +1-{G}_{d}(s)+{G}_{d}(s){G}_{\mathrm{LPF}}(s){K}_{\mathrm{eff}}.
\end{aligned}\tag{13}\label{eq:13}\end{equation}
The closed-loop condition $1+{T}_{v}'(s)=0$ therefore gives

\begin{equation}\begin{aligned}
\Delta_v'(s)&={}sC_f\bigl[sL_f+G_d(s)K_{pi}G_{\mathrm{LPF}}(s)\bigr]\\
&\quad +1-G_d(s)+G_d(s)G_{\mathrm{LPF}}(s)\\
&\quad\times\left[K_{\mathrm{eff}}+\frac{K_{pi}}{sL_{\mathrm{v}}}\right]=0.
\end{aligned}\tag{14}\label{eq:14}\end{equation}

Fig. 6 shows that the reduction retains the resonance-band response while omitting the fundamental PR peak.

\begin{figure}[!t]
\centering
\includegraphics[width=3.173in]{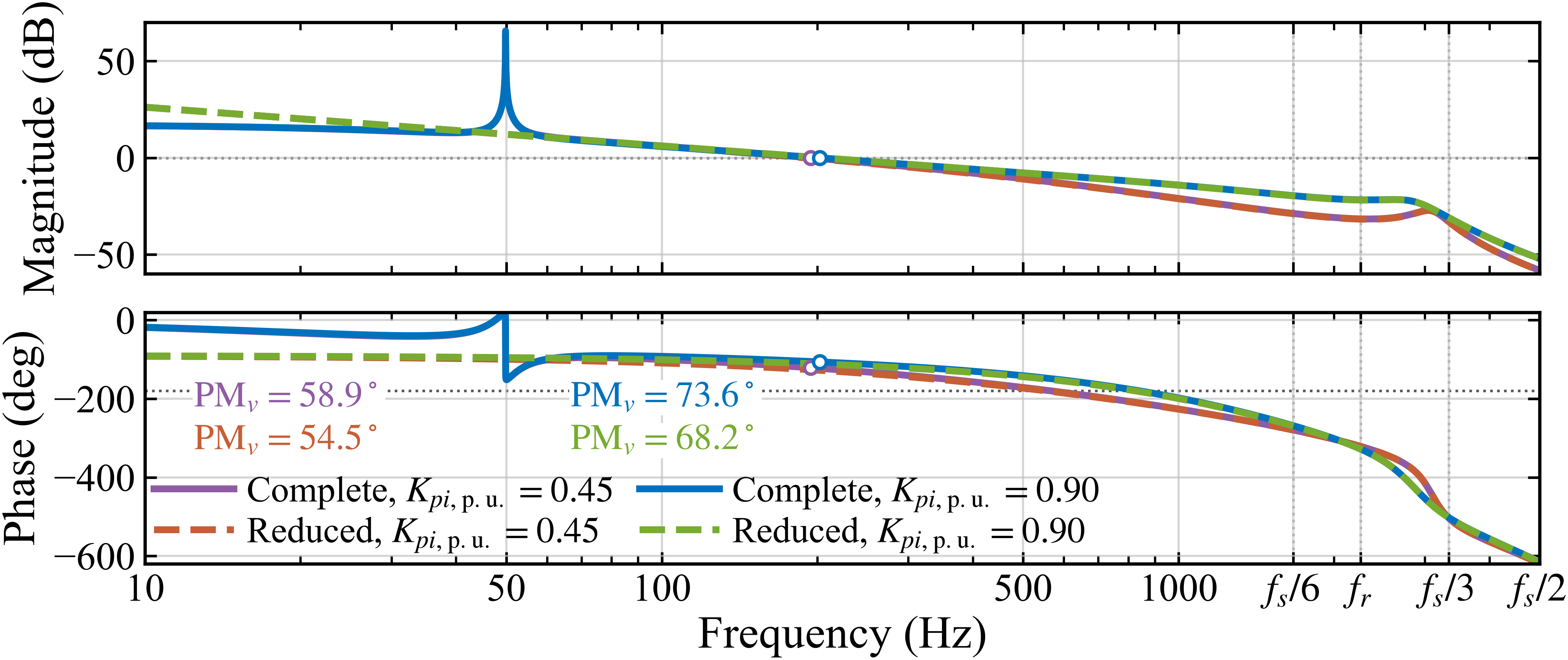}
\par\vspace{-11.9pt}
\caption{Complete and reduced voltage-loop responses, $T_v(s)$ and $T_v^{\prime}(s)$, for the representative filter at $K_{pi,\mathrm{p.u.}}=0.45,0.90$ and $K_{v,\mathrm{p.u.}}=0.70$.}
\label{fig:6}
\end{figure}

\subsection{Imaginary-Axis Crossing and Effective-Gain Boundaries}

Imaginary-axis crossings of (14) define candidate effective-gain boundaries. Pole locations between successive crossings identify the stable interval.

At $s=j\omega$, write $G_d(j\omega)G_{\mathrm{LPF}}(j\omega)=Me^{-j\phi}$, where $M=(1+a_{\mathrm{LPF}}^2)^{-1/2}$ and $\phi=\theta+\tan^{-1}a_{\mathrm{LPF}}$. Defining $B(j\omega)=1-\omega^2L_fC_f-e^{-j\theta}=B_r+jB_i$ and $Q(\omega)=K_{pi}[\omega C_f-1/(\omega L_{\mathrm v})]$, separation into real and imaginary parts gives

\begin{equation}\begin{aligned}
Q\left( \omega \right) &=-\frac{{B}_{r}\sin(\phi) +{B}_{i}\cos(\phi) }{M},\\
{K}_{\mathrm{eff}}(\omega )&=-\frac{{B}_{r}\cos(\phi) -{B}_{i}\sin(\phi) }{M}.
\end{aligned}\tag{15}\label{eq:15}\end{equation}
Appendix C provides the derivation. Solving the first relation gives the crossing frequencies $f_\times=\omega_\times/(2\pi)$; the second gives their effective gains. The search covers $0<f_\times<f_s/2$, including crossings beyond the PVRR.

Fig. 7 applies (15) to the representative filter at $K_{pi,\mathrm{p.u.}}=0.90$ and $f_{\mathrm{LPF}}=0.9f_r$. The crossing residual is $r(\omega)=-Q(\omega)-(B_r\sin(\phi)+B_i\cos(\phi))/M$, normalized by its maximum absolute sampled value over $100\,\mathrm{Hz}\leq f\leq f_s/2$. Its zeros identify candidate crossings; the corresponding gains and pole checks determine the stable interval.

\begin{figure}[!t]
\centering
\includegraphics[width=3.20815in]{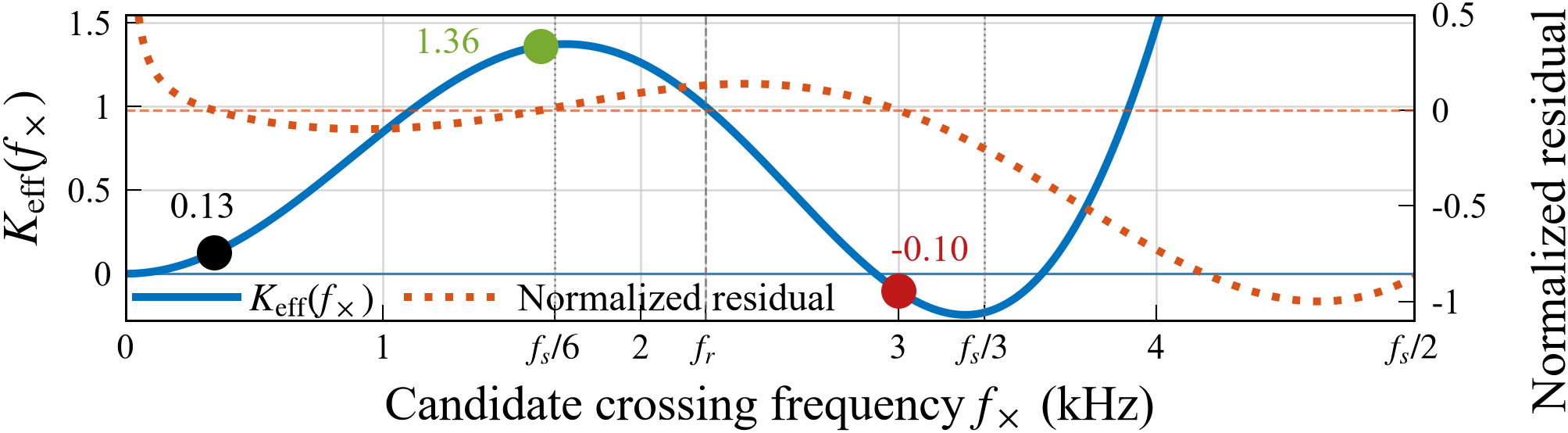}
\par\vspace{-9.8pt}
\caption{Effective gain and normalized crossing residual for the representative filter. Positive-gain crossings locate the lower and upper boundaries.}
\label{fig:7}
\end{figure}

The positive-gain crossings occur at approximately 342~Hz and 1.61~kHz, giving $K_{\mathrm{eff,min}}\approx0.126$ and $K_{\mathrm{eff,max}}\approx1.36$, respectively. The crossing near 2.99~kHz requires negative effective gain and falls outside the positive-$K_v$ design range. Pole analysis identifies the stable reduced-model interval as

\begin{equation}{K}_{\mathrm{eff,min}}<{K}_{\mathrm{eff}}<{K}_{\mathrm{eff,max}}.\tag{16}\label{eq:16}\end{equation}
Insufficient AD destabilizes the voltage-support mode at the lower boundary; excessive AD destabilizes the resonance-band mode at the upper boundary. Their convergence closes the gain window and sets the physical-resonance placement limit. The reduced lower and upper bounds exceed the complete-model values by 21.1\% and 0.79\%, respectively. Lower-boundary sensitivity to omitted VA resistance and residual PR and notch dynamics motivates the complete-model checks in Fig. 9.

\begin{figure}[!t]
\centering
\subfloat[]{\includegraphics[width=1.5656in]{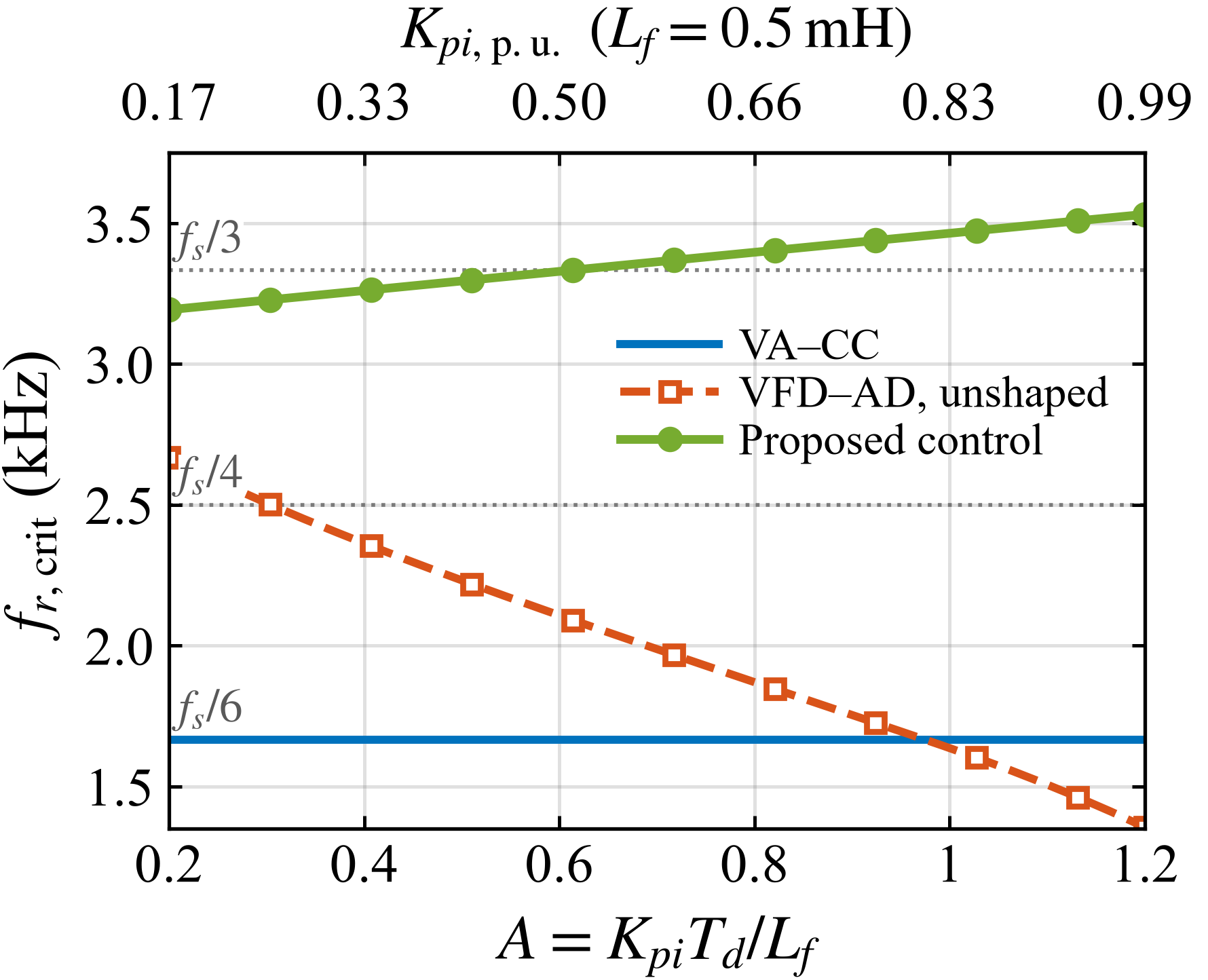}}\hfil\subfloat[]{\includegraphics[width=1.5656in]{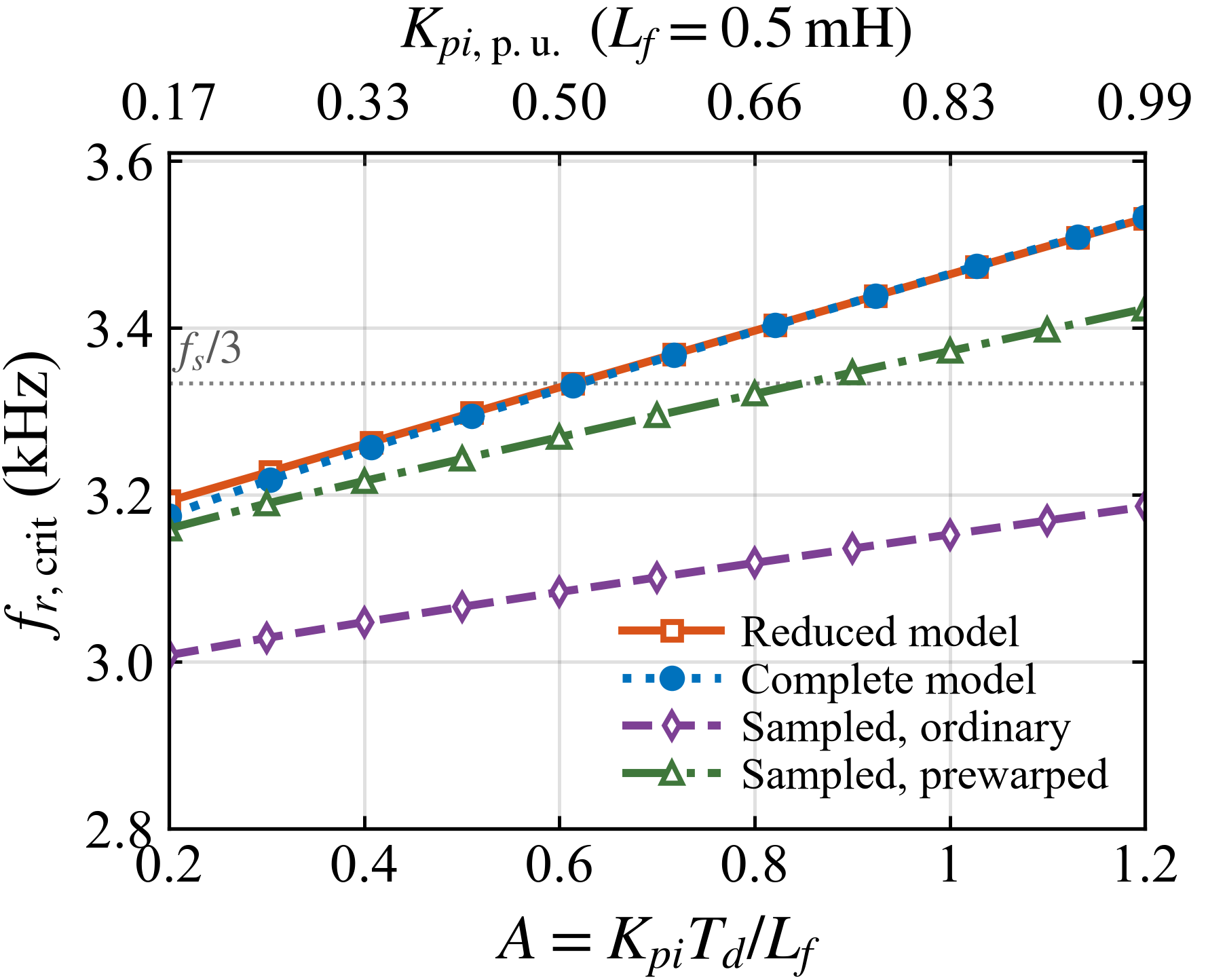}}
\caption{Physical resonance-placement limits with $L_f=0.5\,\mathrm{mH}$ and varying $C_f$: (a) controller comparison and (b) continuous-model and discrete-time implementation predictions with ordinary and prewarped LPFs.}
\label{fig:8}
\end{figure}

\subsection{Control-Shaped Resonance and Placement Limit}

The filter sets the physical resonance $f_r$, while feedback shifts the internal resonance-band mode to $f_r'$. When the resonance-band mode crosses the imaginary axis, $f_r'=f_\times$; within the stable interval, its frequency follows from the closed-loop pole pair. The critical physical resonance occurs when the stable effective-gain boundaries meet:

\begin{equation}{K}_{\mathrm{eff,min}}({f}_{r,\mathrm{crit}})={K}_{\mathrm{eff,max}}({f}_{r,\mathrm{crit}}).\tag{17}\label{eq:17}\end{equation}
Different crossing branches may set the boundaries as the physical resonance increases. Their first closure defines the physical-placement limit. Current-gain dependence is characterized by the dimensionless parameter

\begin{equation}A={K}_{pi}{T}_{d}/{L}_{f}.\tag{18}\label{eq:18}\end{equation}
Here, $A$ is the unshaped proportional-only nominal crossover multiplied by digital control delay; shaping does not change its definition. Fig. 8(a) varies $C_f$ at $L_f=0.5\,\mathrm{mH}$ to locate the first gain-window closure for each $A$. Base VA--CC remains limited to approximately $f_s/6$. Increasing $A$ lowers the unshaped VFD--AD limit from 2.67 to 1.36~kHz, whereas shaping retains 3.17--3.53~kHz. At the smallest $A$, weakened bandwidth separation and resonant-controller residuals constrain controller selection.

Fig. 8(b) shows close agreement between the reduced and complete continuous-model predictions. To quantify discretization effects, a discrete-time implementation model of the LC plant and implemented controller also predicts the resonance-placement limit from gain-window closure (Supplementary Sections~I--III). Its ordinary-Tustin predictions retain the increasing trend but lie 5.3--9.8\% below the complete continuous-model values. Prewarping the LPF reduces this difference to 0.5--3.1\%, identifying frequency warping as a substantial contributor. The experiments use the ordinary LPF without prewarping. At $f_{\mathrm{LPF}}=0.9f_r$, shaping attenuates adverse delayed-current feedback and recovers continuous-model placement limits near the VFD-branch resistance boundary:

\begin{equation}{f}_{r,\mathrm{crit}}\approx1/(2{T}_{d})={f}_{s}/3.\tag{19}\label{eq:19}\end{equation}
Feedback shifts the internal mode, separating the branch-resistance, PVRR, and gain-window limits. Equation (19) provides a nominal continuous-model placement guide; discretization shifts its predicted limit as quantified in Fig. 8(b).

\subsection{Design Procedure and Operating-Point Selection}

Fig. 9 coordinates design at $f_{\mathrm{LPF}}=0.9f_r$. Equations (11), (15)--(17), and (10) initialize current gain, screen gain and placement, and initialize $K_v$ from voltage bandwidth. The complete analytical model then checks internal stability, voltage response, and damping. Grid-connected designs additionally require impedance-based stability over the intended grid range (Section IV-C2); failed checks require retuning.

\begin{figure}[!t]
\centering
\includegraphics[width=3.20815in]{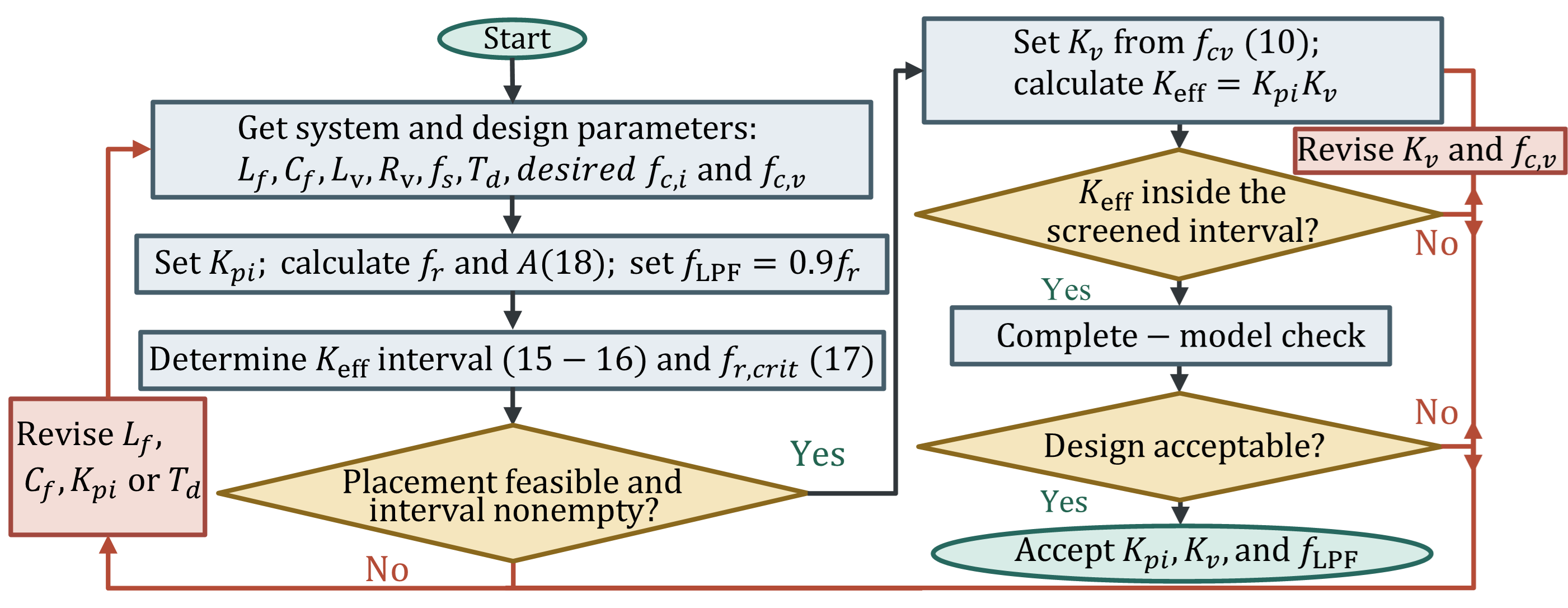}
\par\vspace{-5.5pt}
\caption{Coordinated filter selection, gain tuning, and model verification.}
\label{fig:9}
\end{figure}

Table II compares operating points A--C\ensuremath{^{\prime}} of the proposed controller with unshaped U under no load at $K_{v,\mathrm{p.u.}}=0.70$. For A--C\ensuremath{^{\prime}}, $f_{c,v}=194$--$211$~Hz lies within 5\% of the 204-Hz estimate in (10). Nominal current metrics use (11); complete-model voltage metrics use \eqref{eq:B3} and \eqref{eq:B4}. A--C compare LC reduction at comparable crossovers; U and C\ensuremath{^{\prime}} compare current gain with the filter and AD gain fixed.

\begin{table}[!t]

\caption{No-Load Operating Points and Loop Metrics ($f_{c,i}$ and $\mathrm{PM}_i$: Nominal)}
\label{tab:2}
\centering\footnotesize
\setlength{\tabcolsep}{1.7pt}
\renewcommand{\arraystretch}{1.0}
\begin{tabular}{@{}lccccccc@{}}
\toprule
Point & \shortstack{$L_f/C_f$\\(mH/\textmu F)} & ${f}_{r}$ (Hz) & ${K}_{pi,\mathrm{p.u.}}$ & ${f}_{c,i}$ (Hz) & ${f}_{c,v}$ (Hz) & \shortstack{$\mathrm{PM}_i/\mathrm{PM}_v$\\(${}^\circ$)} & $\zeta_{\mathrm{min}}$ \\
\midrule
A & 1.0/30 & 919 & 0.90 & 501 & 211 & 30.5/64.0 & 0.176 \\
B & 1.0/10 & 1592 & 0.90 & 544 & 208 & 38.6/72.8 & 0.404 \\
C & 0.5/10 & 2251 & 0.45 & 563 & 194 & 41.8/58.9 & 0.0715 \\
C\ensuremath{^{\prime}} & 0.5/10 & 2251 & 0.90 & 1035 & 202 & 6.45/73.6 & 0.135 \\
U & 0.5/10 & 2251 & 0.30 & 387 & 177 & 64.1/50.4 & 0.00421 \\
\bottomrule
\end{tabular}
\end{table}

Fig. 10(a) compares complete- and reduced-model stability regions. Constant-$K_{v,\mathrm{p.u.}}$ lines locate AD gains. Excluding 10\% of the nominal internal-stability window width at each boundary retains the middle 80\% as the screened interval. This clearance is defined in gain space; component tolerances and grid-connected stability require separate checks.

\begin{figure}[!t]
\centering
\subfloat[]{\includegraphics[width=3.377in]{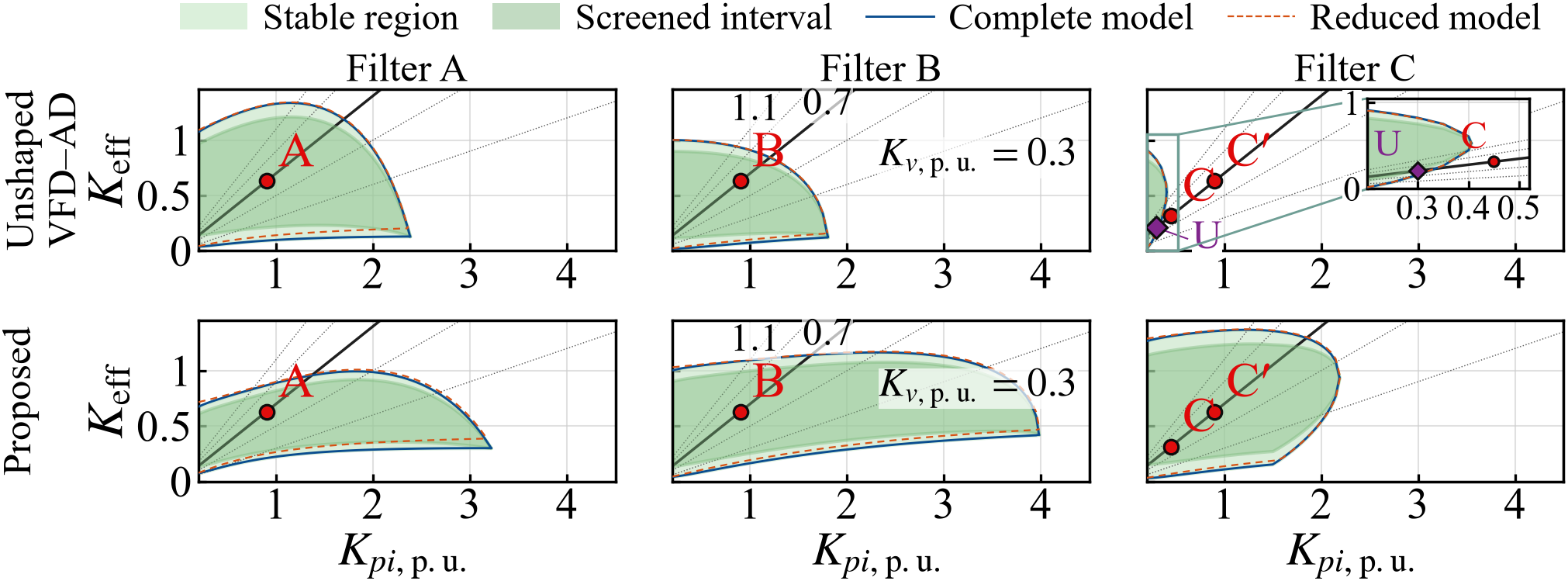}}
\par\vspace{1pt}
\subfloat[]{\includegraphics[width=3.377in]{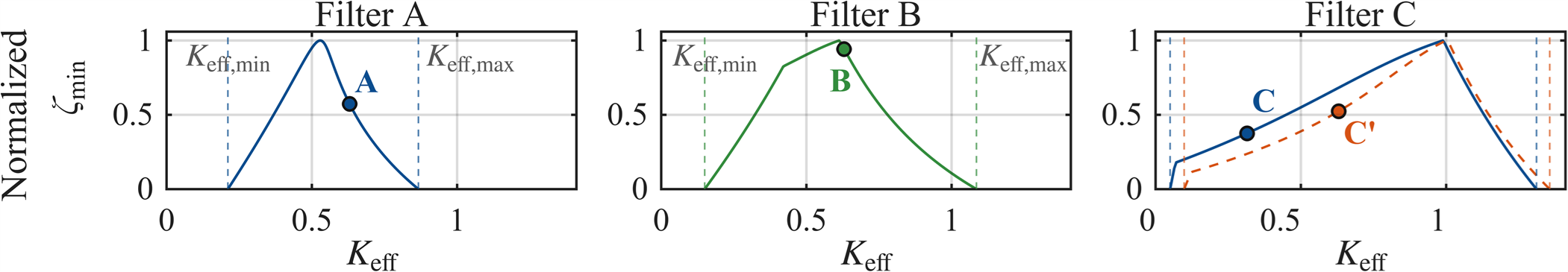}}
\caption{Gain-window stability and damping: (a) unshaped VFD--AD and shaped regions for filters A--C, with AD-gain lines and points (purple diamond: U); (b) normalized minimum modal damping for Cases A--C\ensuremath{^{\prime}}. Solid/dashed: complete/reduced boundaries; darker shading: screened interval.}
\label{fig:10}
\end{figure}

The highest-resonance filter exhibits the strongest unshaped-window contraction, excluding Cases C and C\ensuremath{^{\prime}}. Shaping accommodates both, expanding filter and bandwidth choices. AD retuning cannot reopen a closed window; current gain, filter parameters, or frequency shaping must change.

Stable operating points can have substantially different resonance damping. After selecting $K_v$ for voltage bandwidth within the screened interval, the least-damped oscillatory mode is quantified from closed-loop poles $p_m$ between 100~Hz and $f_s/2$:

\begin{equation}{\zeta }_{\mathrm{min}}({K}_{\mathrm{eff}})={\min }_{m}\left[ \frac{-\Re {{p}_{m}({K}_{\mathrm{eff}})}}{\left| {p}_{m}({K}_{\mathrm{eff}})\right| }\right] .\tag{20}\label{eq:20}\end{equation}
Fig. 10(b) normalizes each curve by its maximum for within-case comparison. Table II reports unnormalized values from~(20).

From Case C to C\ensuremath{^{\prime}}, nominal $\mathrm{PM}_i$ falls from $41.8^\circ$ to $6.45^\circ$ while $\zeta_{\mathrm{min}}$ rises from 0.0715 to 0.135. This is consistent with Fig. 5: (11) omits VFD--AD modification of the current-loop plant. At voltage crossover, dominant AD support approximately cancels $G_dG_i$ in (12), preserving the higher $\mathrm{PM}_v$. Thus, nominal $\mathrm{PM}_i$ describes a tuning tradeoff, while the actual loop gains and complete-model poles establish stability and damping.

\setcounter{figure}{10}
\begin{figure}[!t]
\centering
\includegraphics[width=2.8804in]{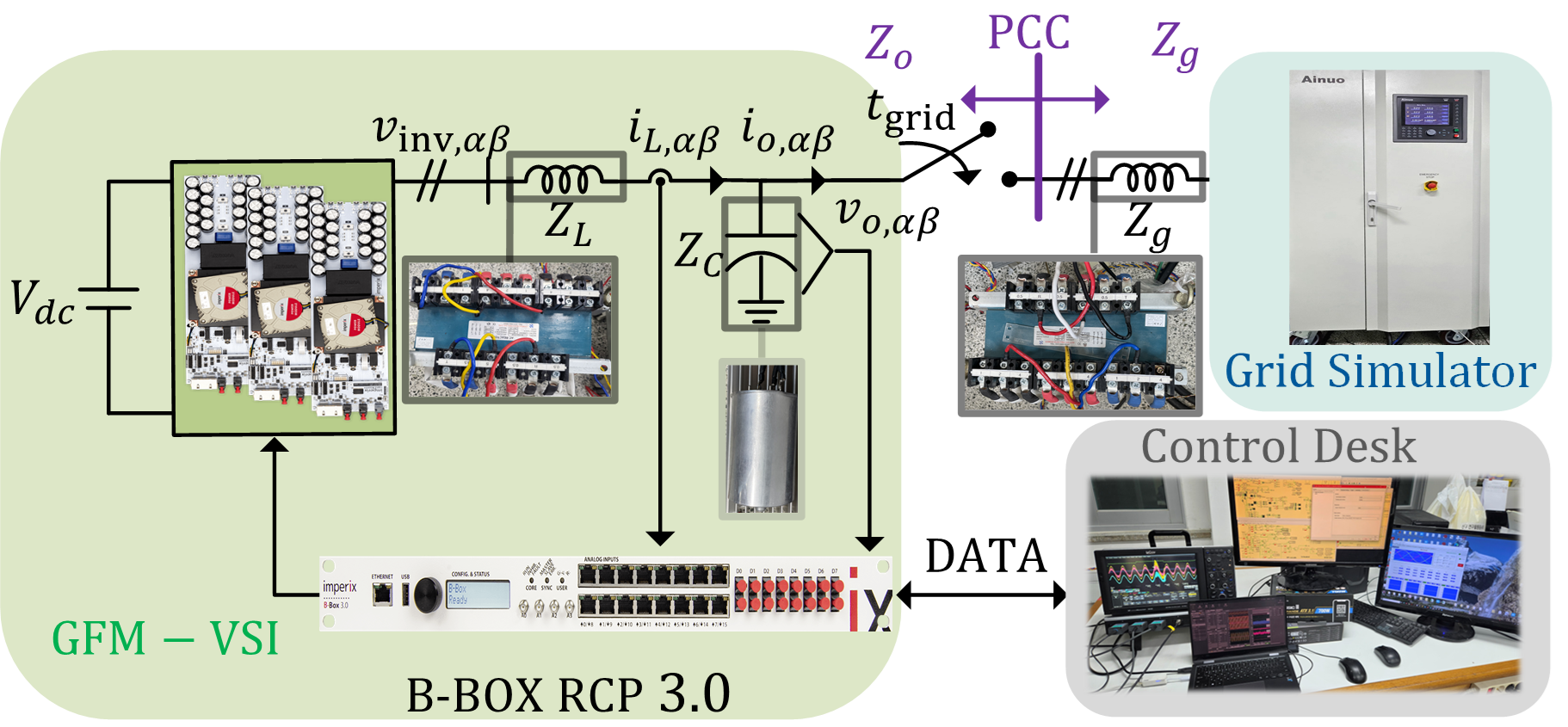}
\par\vspace{-5.5pt}
\caption{Unloaded and grid-connected experimental platform. PCC: point of common coupling.}
\label{fig:11}
\end{figure}

\setcounter{figure}{11}
\begin{figure*}[!t]
\centering
\subfloat[]{\includegraphics[width=2.29in]{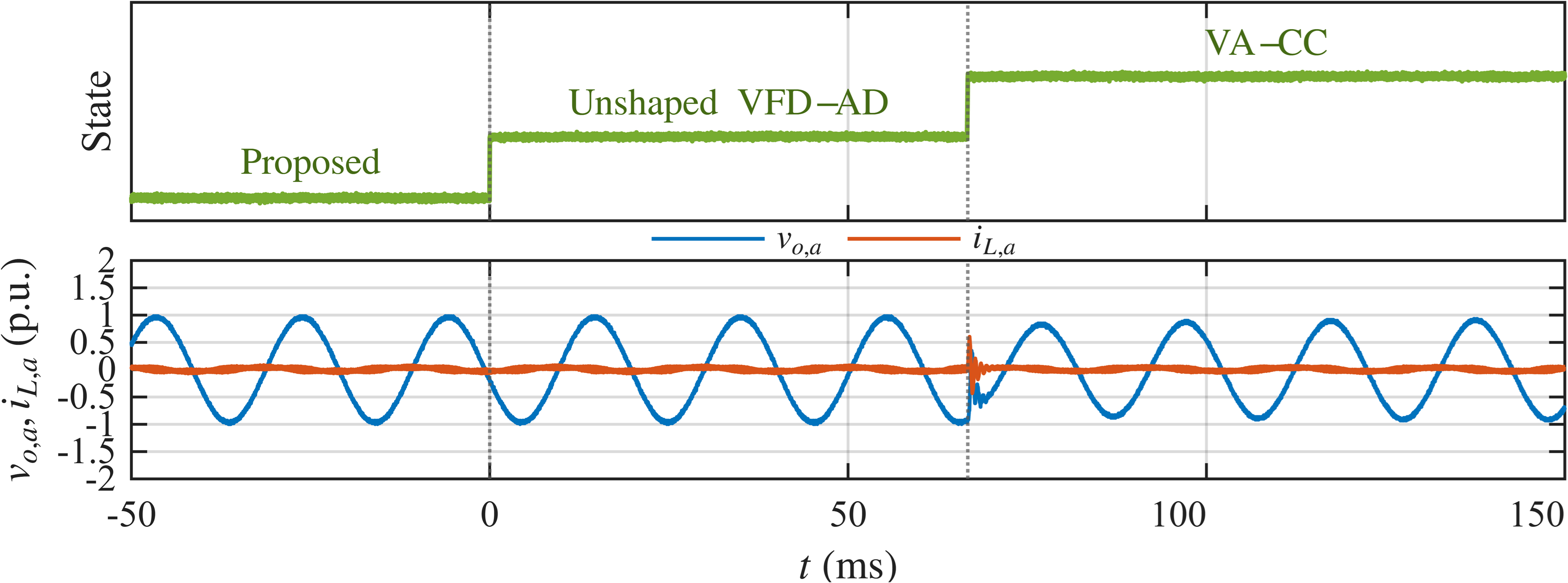}}\hfil\subfloat[]{\includegraphics[width=2.29in]{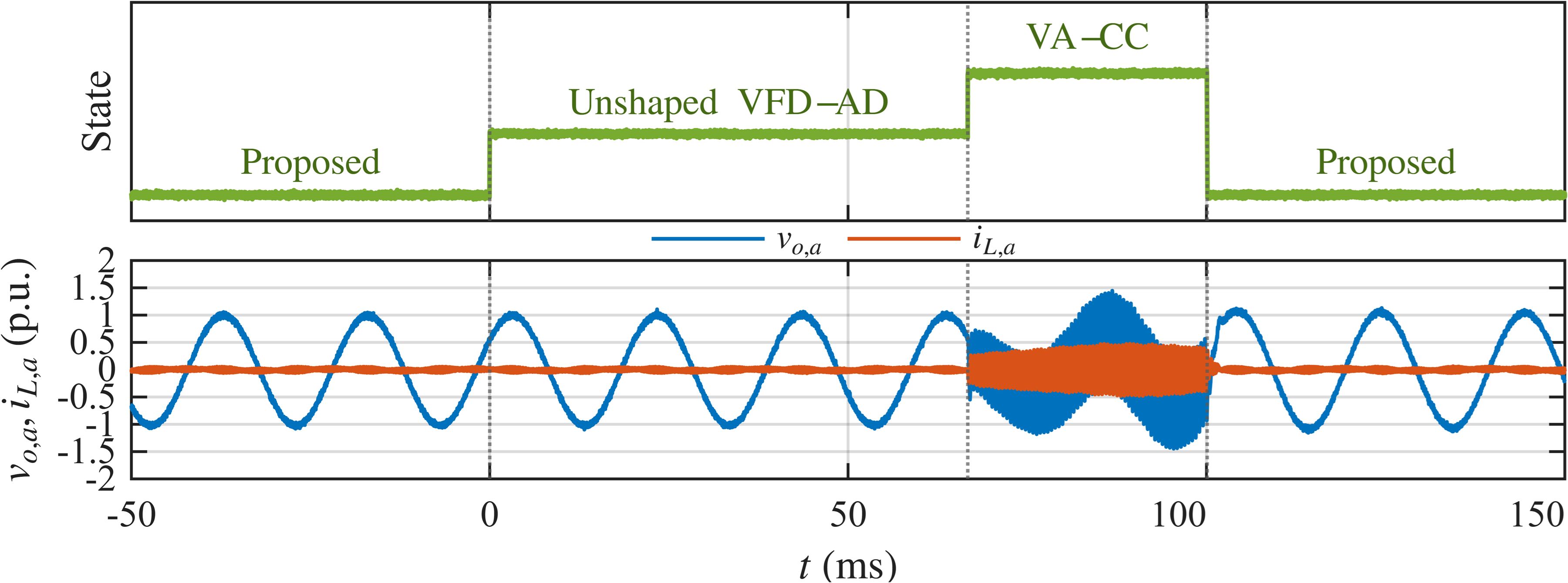}}\hfil\subfloat[]{\includegraphics[width=2.29in]{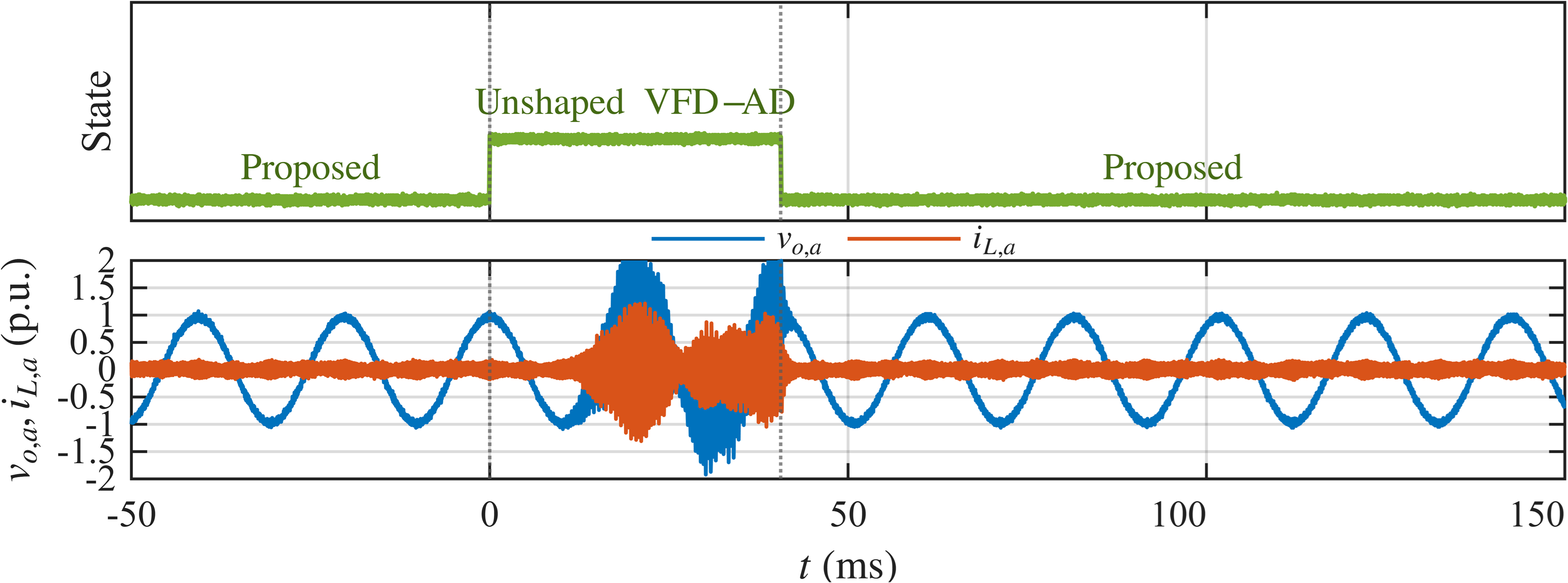}}
\caption{Unloaded controller transitions: (a) Case A, (b) Case B, and (c) Case C. The proposed controller is restored after instability develops in (b) and (c).}
\label{fig:12}
\end{figure*}

\setcounter{figure}{12}

\begin{figure}[!t]
\centering
\subfloat[]{\includegraphics[width=3.20815in]{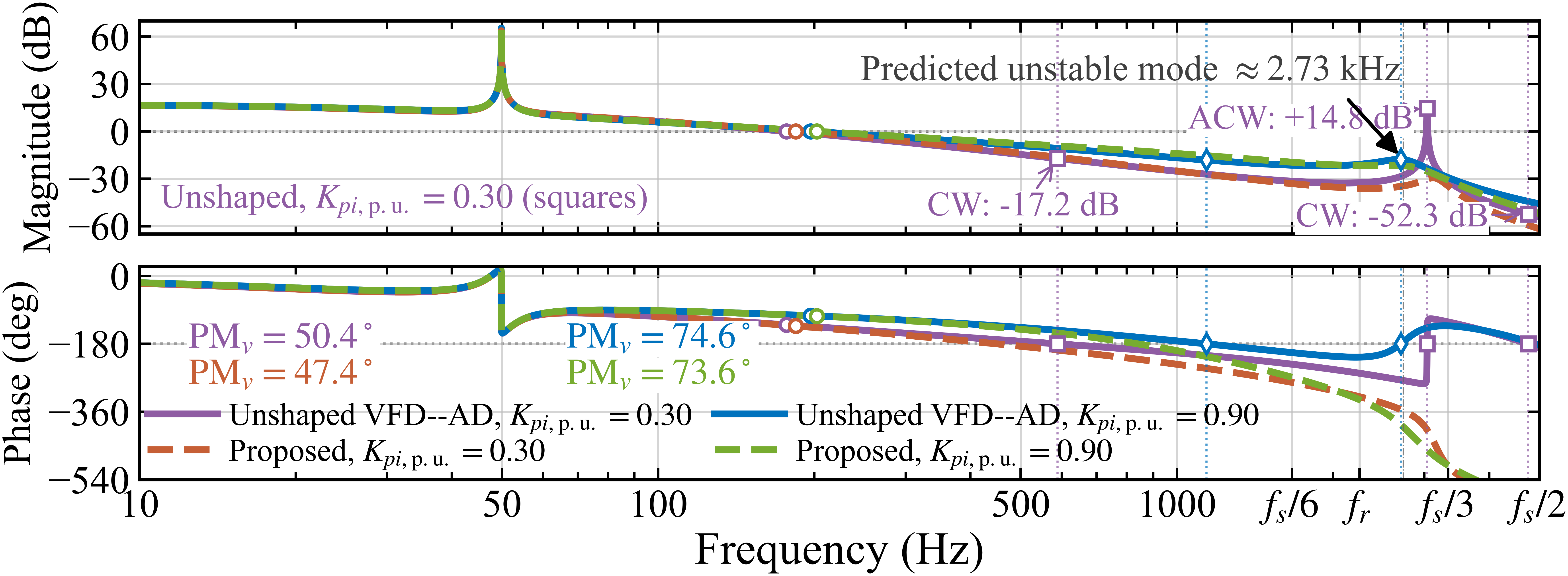}}
\par\vspace{1pt}
\subfloat[]{\includegraphics[width=3.20815in]{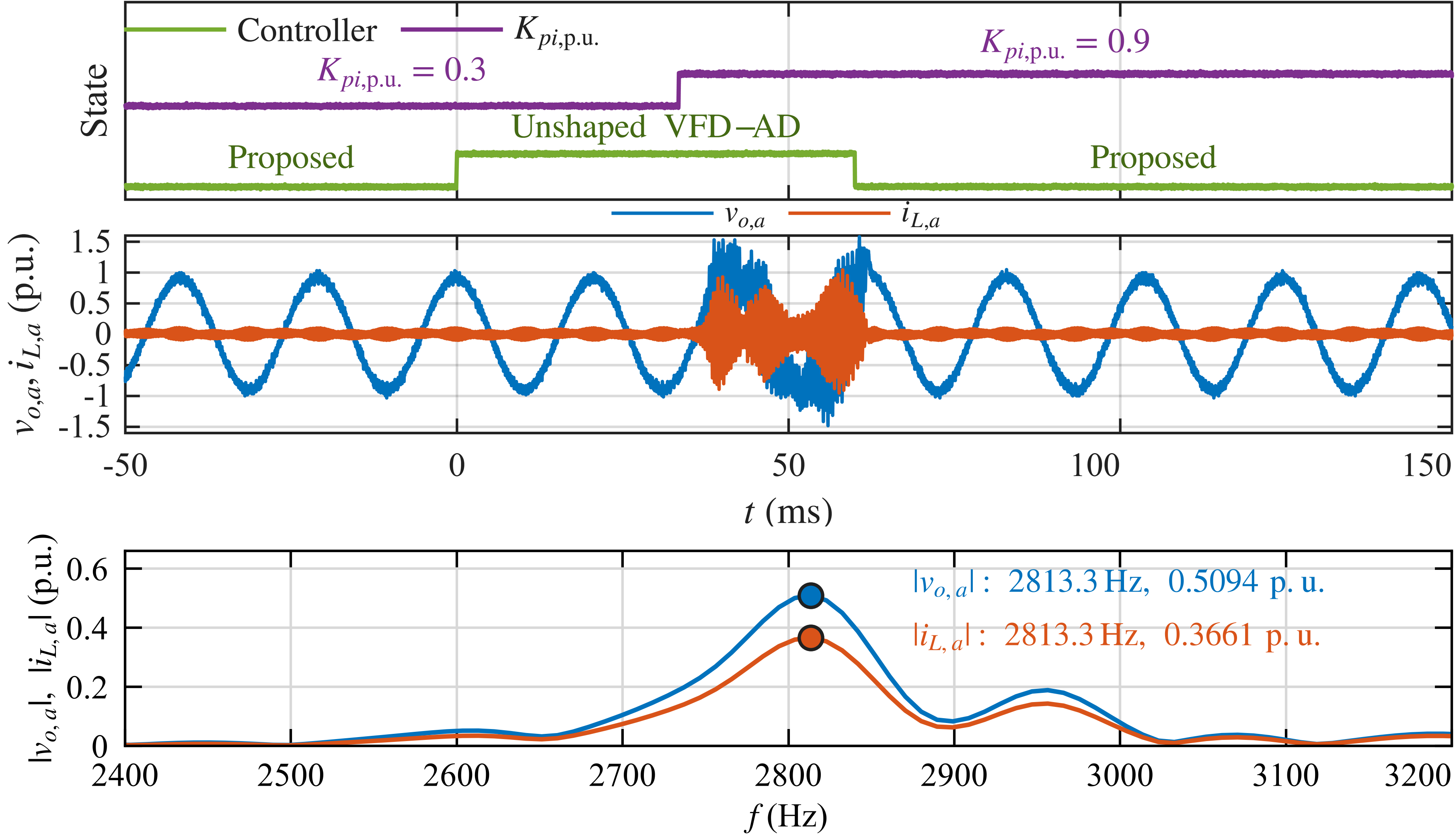}}
\caption{Current-gain-dependent instability and recovery with the proposed controller: (a) complete voltage-loop responses and (b) unloaded transitions and measured FFT results.}
\label{fig:13}
\end{figure}

\begin{figure}[t]
\centering
\subfloat[]{\includegraphics[width=3.20815in]{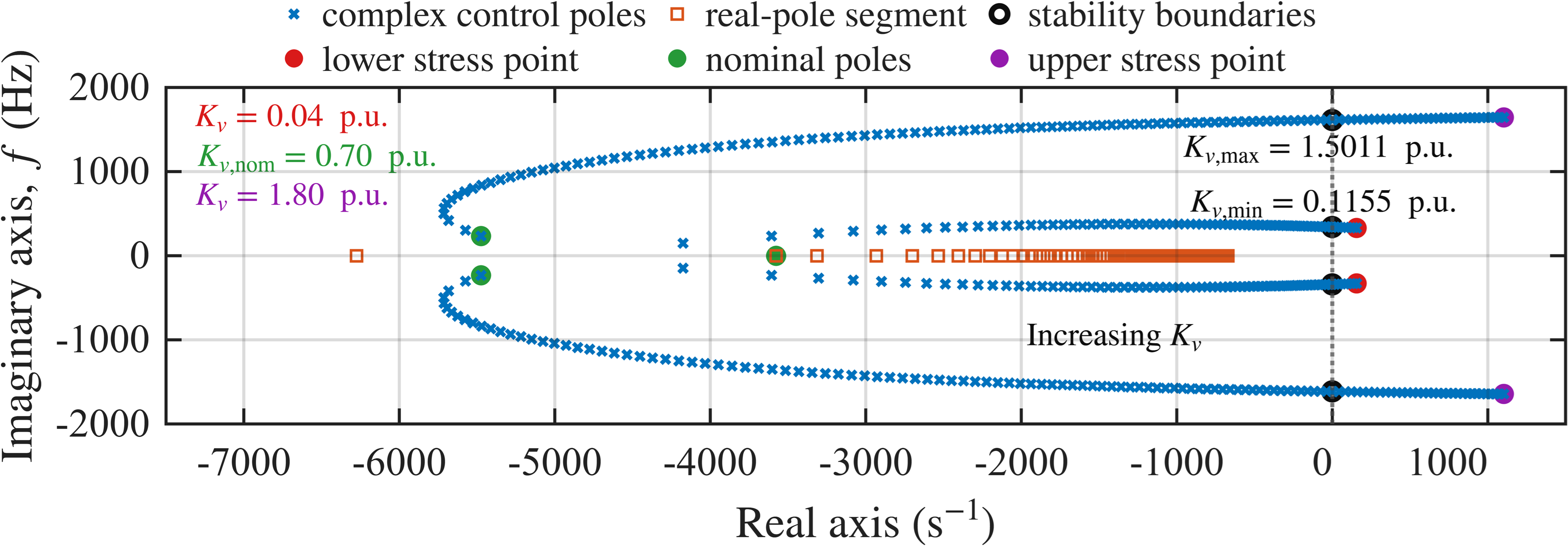}}
\par\vspace{1pt}
\subfloat[]{\includegraphics[width=3.20815in]{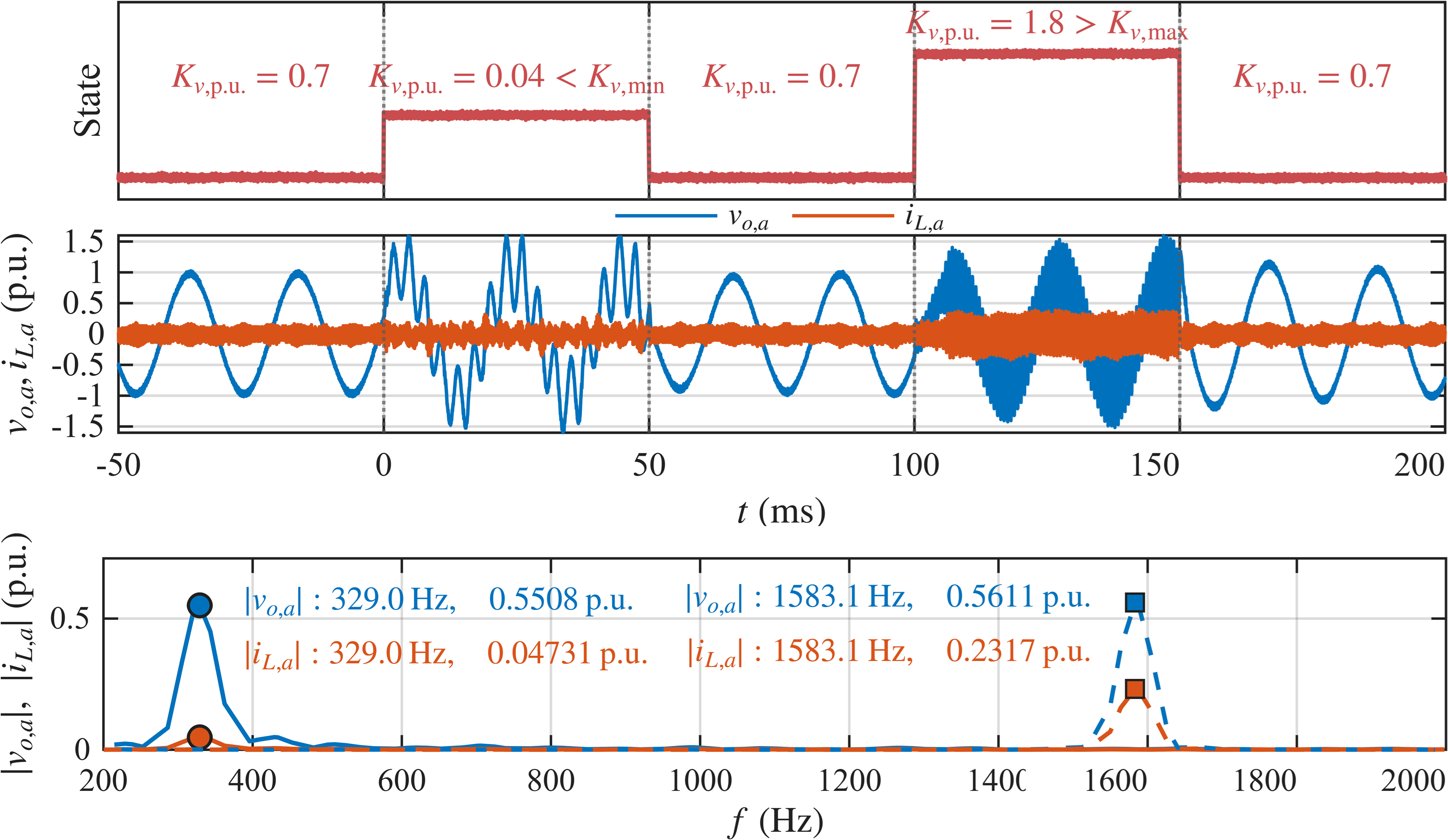}}
\caption{Voltage-loop gain window: (a) complete-model pole locus; (b) gain transitions and FFT results (solid/dashed: lower/upper gains). Pole migration is animated in Supplementary Video S1.}
\label{fig:14}
\end{figure}

\section{Experimental Verification and Discussion}

\subsection{Experimental Setup}

Fig. 11 shows the 3-kVA prototype: Imperix PEB 8038 power modules, an LC filter, an Imperix B-Box RCP 3.0 controller, an IT6018T bidirectional dc supply, an ANBGS grid simulator, and a host computer. Tables I and II give parameters. The experiments use 10-kHz SSSU PWM and an ordinary Tustin realization of the proportional-path LPF without prewarping.

In unloaded tests, the PSC remains enabled, the RPC is disabled, and the IVS magnitude is fixed at its nominal value. Grid-connected tests enable both power controllers and inject 0.5 p.u. active power into inductive grids.

Waveforms use the peak phase bases in Section II-A. Fast Fourier transform (FFT) peaks are the largest bins within selected bands after mean removal and periodic Hann windowing; comparisons use predictions at the applied gains, separately from boundary-crossing frequencies. Total harmonic distortion (THD) uses one fundamental period at the initial nominal gain. It is the residual-to-fundamental RMS ratio after removing DC and the least-squares-fitted fundamental, retaining captured switching ripple (Supplementary Section~IV).

\subsection{No-Load Internal Stability and Resonance Placement}

\subsubsection{Physical-Resonance Placement}

Cases A--C test LC-product reduction at comparable crossovers (Table II). Complete-model poles predict all three controllers stable in Case A; in Case B, base VA--CC becomes unstable while VFD--AD remains stable; in Case C, unshaped VFD--AD also loses stability, consistent with Fig. 10(a). Fig. 12 tests this progression with each filter and its current-controller gains unchanged during transitions.

The measurements follow these predictions: A remains stable, whereas B and C oscillate under their respective baselines. Restoring the proposed controller suppresses the oscillations. Shaping supports an 83.3\% reduction in $L_fC_f$ from A to C, with stable operation at 2.251~kHz and comparable voltage and nominal current crossovers.

\subsubsection{Proportional-Gain Coupling and Controller Recovery}

For the Case C filter at $K_{v,\mathrm{p.u.}}=0.70$, Fig. 13(a) compares $T_v$ at $K_{pi,\mathrm{p.u.}}=0.30,0.90$. At 0.30, clockwise (CW) $-180^\circ$ crossings lie below unity and the anticlockwise (ACW) crossing above it. Shaping suppresses this resonance peak while largely preserving voltage crossover. At 0.90, unshaped $T_v$ inherits an RHP pole pair from the current-loop closure in Fig. 5. Both resonance-band $-180^\circ$ crossings lie below unity: the curve has zero net encirclements instead of the two anticlockwise encirclements required by its RHP-pole count. These crossings violate the paired gain-margin condition \cite{Liao2020PassivityVoltageControllers,Woo2026DifferentialCapacitorCurrent}, despite a positive voltage-crossover phase margin. Shaping removes the RHP pair, restoring zero net encirclement (Supplementary Fig.~S2). Complete-model poles predict a 2.730-kHz unstable mode without shaping.

Consistently, Fig. 13(b) remains stable without shaping at $K_{pi,\mathrm{p.u.}}=0.30$ but oscillates when gain rises to 0.90. Restoring shaping suppresses the oscillation without reducing gain. Voltage and current FFT peaks are both 2.813~kHz; the voltage peak exceeds the prediction by 3.06\%.

Point U is internally stable but weakly damped and outside the screened interval; C\ensuremath{^{\prime}} passes screening with greater damping (Table II). Their nominal current crossovers are 387~Hz and 1.03~kHz; the shaped value is 2.68 times the unshaped value with the same filter and AD gain. Together with Fig. 12, this demonstrates LC-product reduction at comparable crossovers or higher current gain with the same filter, illustrating feasibility rather than optimality.

\subsubsection{Two-Sided Voltage-Loop Gain Window}

AD must support the voltage loop while preserving resonance damping. For Case C\ensuremath{^{\prime}}, Fig. 14(a) places the nominal $K_{v,\mathrm{p.u.}}=0.70$ within $0.1155<K_{v,\mathrm{p.u.}}<1.5011$. The voltage-support and resonance-band modes cross into the RHP at the lower and upper boundaries at 342.7~Hz and 1.611~kHz, respectively. At the applied gains of 0.04 and 1.80, their predicted unstable frequencies are 330.5~Hz and 1.643~kHz. Thus, stable current-loop closure alone is insufficient: the lower applied gain loses voltage-loop support, whereas the upper gain reintroduces an RHP pair.

Fig. 14(b) confirms both mechanisms: measured peaks at 329.0~Hz and 1.583~kHz differ from predictions by 0.44\% and 3.62\%. Restoring nominal gain suppresses both oscillations. Supplementary Section~VI tests the delay dependence of these limits using double-sampled double-update operation.

\subsection{Grid-Connected Stability}

The grid-connected tests examine weak-grid damping recovery and the additional external-stability constraints imposed by grid impedance.

\subsubsection{Damping Recovery Under Weak-Grid Operation}

At a short-circuit ratio (SCR) of 2.1, grid-connected poles predict the same controller ordering: Case B is unstable under base VA--CC but stable under VFD--AD, whereas Case C requires the proposed shaping. Fig. 15 tests these predictions through controller transitions.

\begin{figure}[!t]
\centering
\subfloat[]{\includegraphics[width=3.20815in]{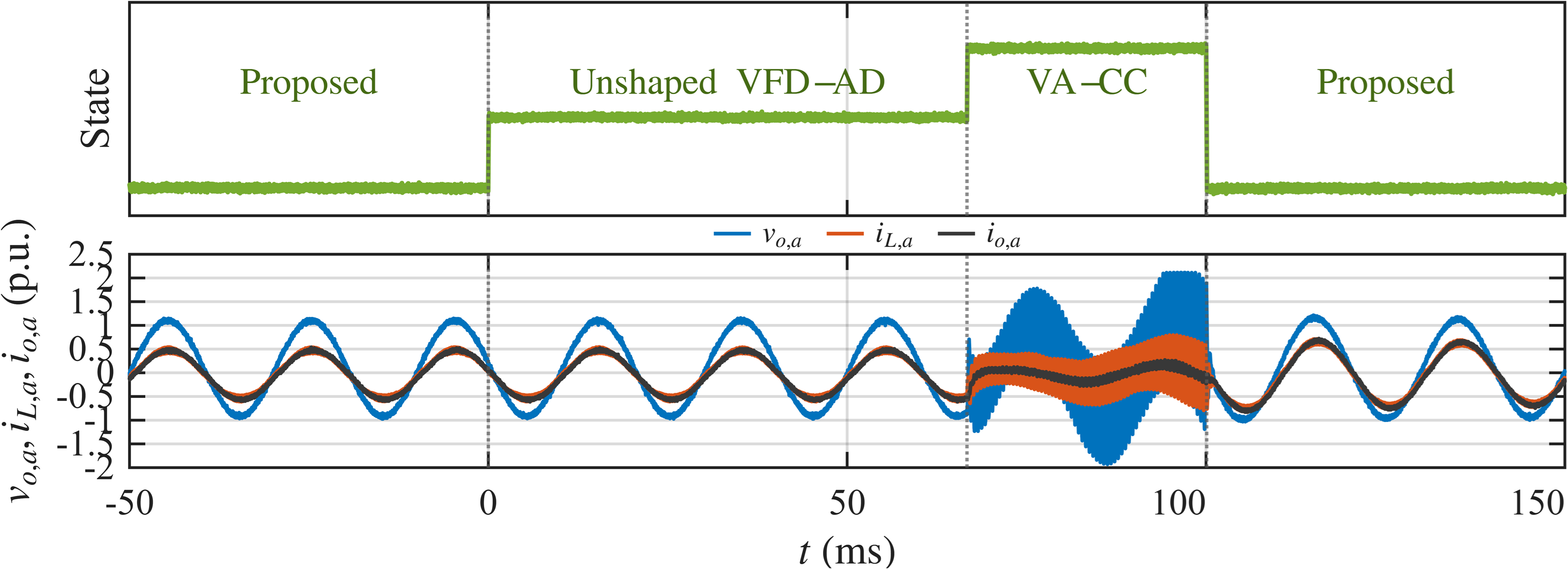}}
\par\vspace{1pt}
\subfloat[]{\includegraphics[width=3.20815in]{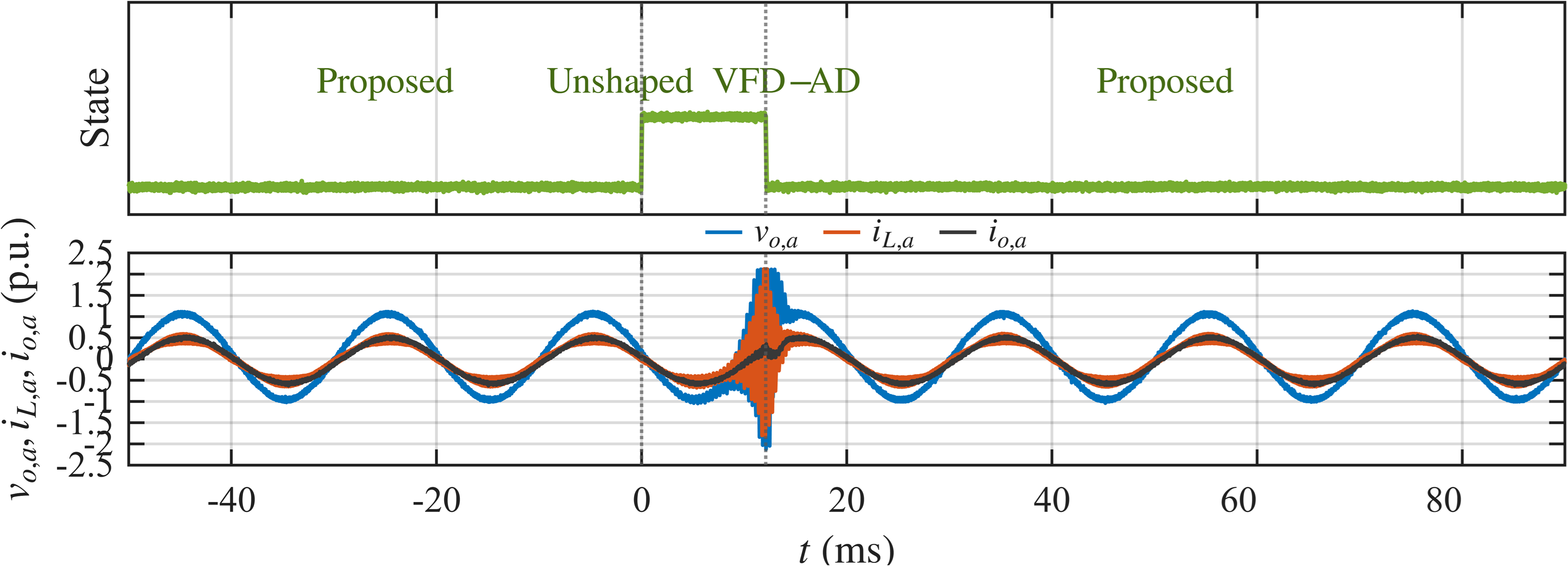}}
\caption{Controller transitions at $\mathrm{SCR}=2.1$: (a) Case B and (b) Case C. The proposed controller is restored after instability develops.}
\label{fig:15}
\end{figure}

Fig. 15(a) reproduces the predicted Case B contrast. In Fig. 15(b), bypassing shaping destabilizes C and restoring it suppresses the oscillation, corroborating damping recovery under the tested weak-grid connection with both power controllers enabled.

\subsubsection{Output Impedance and Grid-Interaction Limits}

External stability concerns interaction with a load or grid and is not ensured by internal stability. The same SISO model assesses grid interaction for $f\geq2f_{\mathrm{o}}$ (100~Hz) \cite{Liao2020PassivityVoltageControllers}, \cite{He2025FeedforwardDamping}. Setting the voltage-reference perturbation to zero and perturbing current at the capacitor node gives

\begin{equation}{Z}_{o}(s)=\frac{s{L}_{f}+{G}_{d}(s){G}_{i}(s)}{{D}_{v}(s)+{G}_{d}(s){G}_{i}(s){Y}_{\mathrm{v}}(s)}.\tag{21}\label{eq:21}\end{equation}

\begin{figure}[!t]
\centering
\subfloat[]{\includegraphics[width=3.20815in]{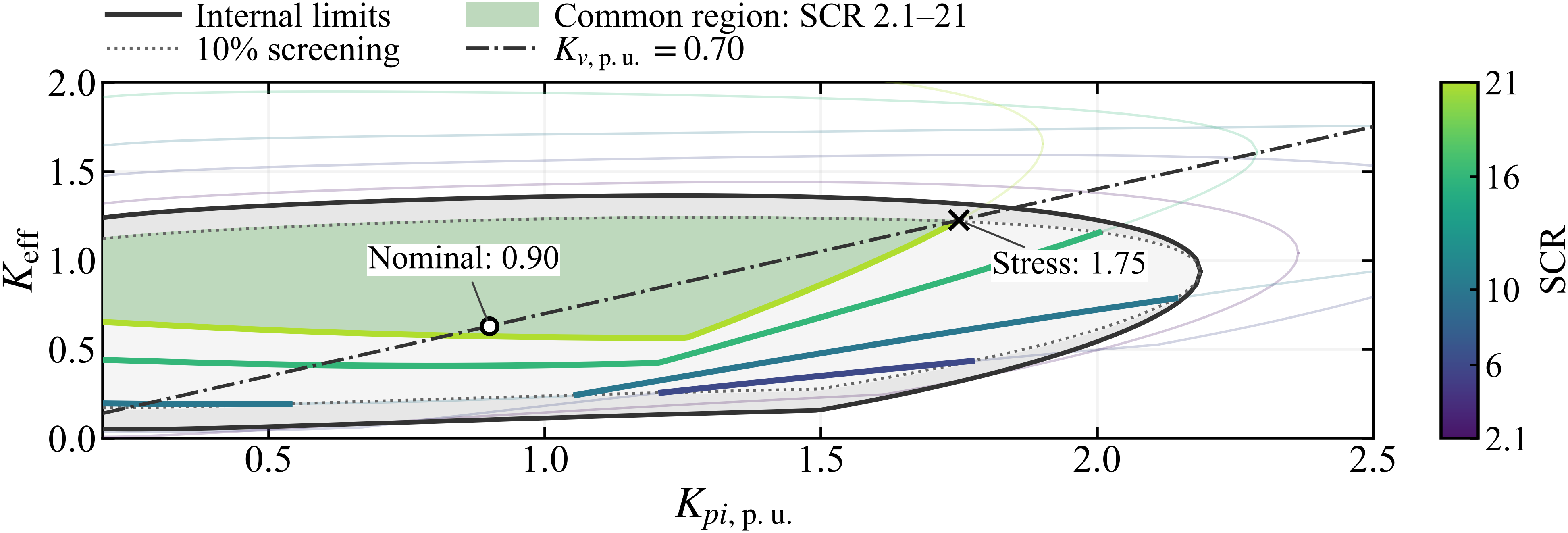}}
\par\vspace{1pt}
\subfloat[]{\includegraphics[width=3.20815in]{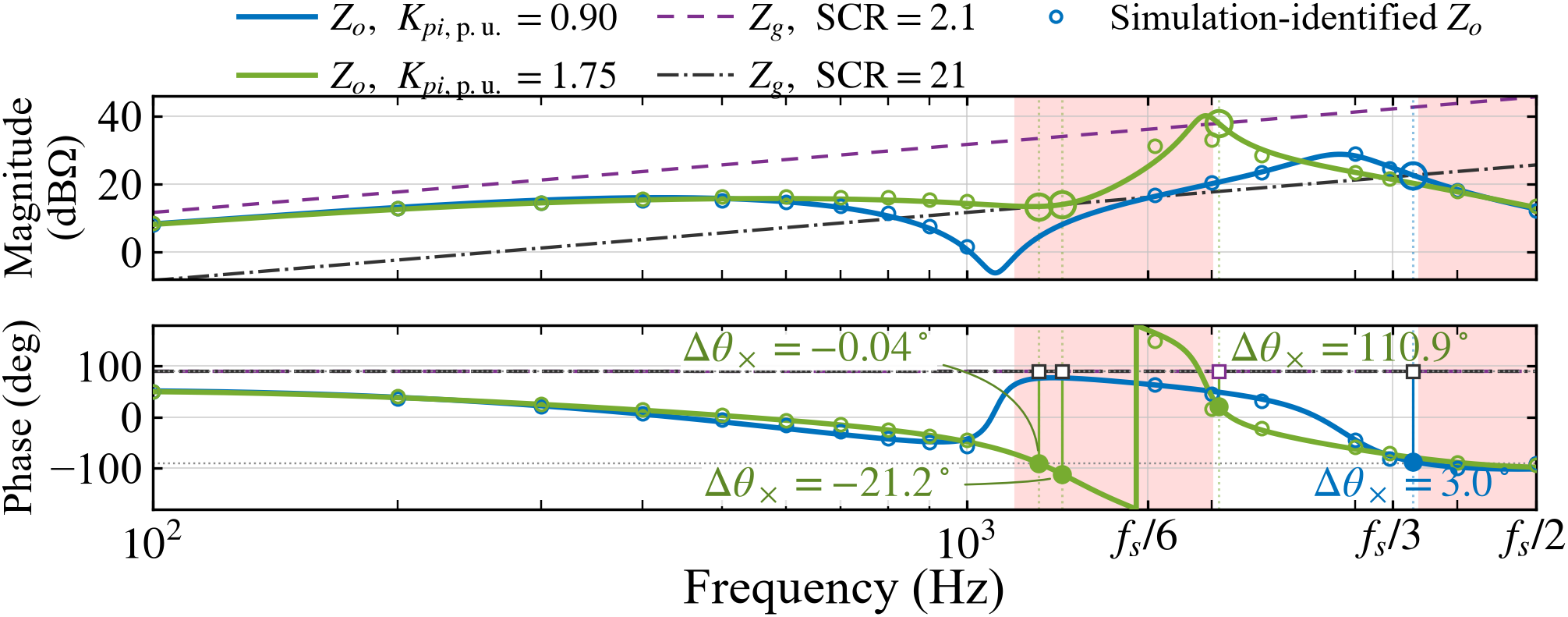}}
\par\vspace{-0.7pt}
\caption{Grid interaction for the Case C\ensuremath{^{\prime}} filter: (a) effective-gain regions with SCR-coded grid boundaries (green: common region over evaluated SCR = 2.1--21); (b) analytical (lines) and simulation-identified (circles) impedances.}
\label{fig:16}
\end{figure}
Appendix B derives (21), with $D_v(s)$ defined in (12). The predominantly inductive test grids are approximated by $Z_g(s)=sL_g$, where the grid inductance is related to SCR by

\begin{equation}{L}_{g}={Z}_{\mathrm{b}}/(\mathrm{SCR}\,{\omega}_{o}).\tag{22}\label{eq:22}\end{equation}
For grid interaction, the Nyquist criterion applies to $Z_o/Z_g$, with characteristic condition $1+Z_o(s)/Z_g(s)=0$. Substituting (21), with $\Delta(s)=D_v(s)+G_d(s)G_i(s)Y_{\mathrm v}(s)$, gives

\begin{equation}{Z}_{g}(s)\Delta (s)+s{L}_{f}+{G}_{d}(s){G}_{i}(s)=0.\tag{23}\label{eq:23}\end{equation}
Dividing (23) by $sL_g$ gives the grid contribution $L_f/L_g+G_dG_i/(sL_g)$. Increasing SCR reduces $L_g$, strengthening both terms; the latter carries current-controller phase and digital control delay. Consequently, grid interaction can impose gain limits within the internal stability window.

Fig. 16(a) extends Fig. 10 to grid connection with $K_v=K_{\mathrm{eff}}/K_{pi}$. SCR shifts both bounds; some grid-stable points are internally unstable. The green intersection retains the internally screened and grid-stable regions over the evaluated SCR range of 2.1--21, preserving inner-loop stability if the grid disconnects. $K_v$ from (10) selects a constant-$K_v$ current-gain path. At $K_{v,\mathrm{p.u.}}=0.70$ and SCR = 21, the continuous interval $0.8241<K_{pi,\mathrm{p.u.}}<1.7498$ excludes 1.75. The discrete-time implementation predicts 0.808--1.551 p.u., with lower/upper bounds 2.0\%/11.4\% below the continuous predictions. Both models classify 0.90 as grid-stable and 1.75 as grid-unstable while both remain internally stable. Thus, discretization shifts boundaries without removing the grid-dependent constraint.

Fig. 16(b) compares analytical impedances with PLECS-identified points from PCC perturbations while the controller executes as the discrete subsystem used for code generation. Nominal-gain $Z_o$ is mostly passive over the plotted band; the stress gain introduces a nonpassive pocket around $f_s/6$. At magnitude intersections $\omega_\times$, define $\Delta\theta_\times=180^\circ-|\angle Z_o-\angle Z_g|$. The criterion uses the RHP-pole count and net directional count of opposing-phase crossings where $|Z_o/Z_g|>1$ \cite{Liao2020ImpedanceRHPCriterion}. At the stress gain and SCR = 21, both intersections (1.226 and 1.308~kHz) have negative phase separations. The criterion confirms grid-interaction instability; poles of (23) locate the unstable mode at 1.226~kHz near the first intersection. Internal stability is retained, and SCR = 2.1 remains stable. Supplementary Fig.~S3 gives the impedance-ratio Nyquist plots.

\begin{figure}[!t]
\centering
\subfloat[]{\includegraphics[width=3.20815in]{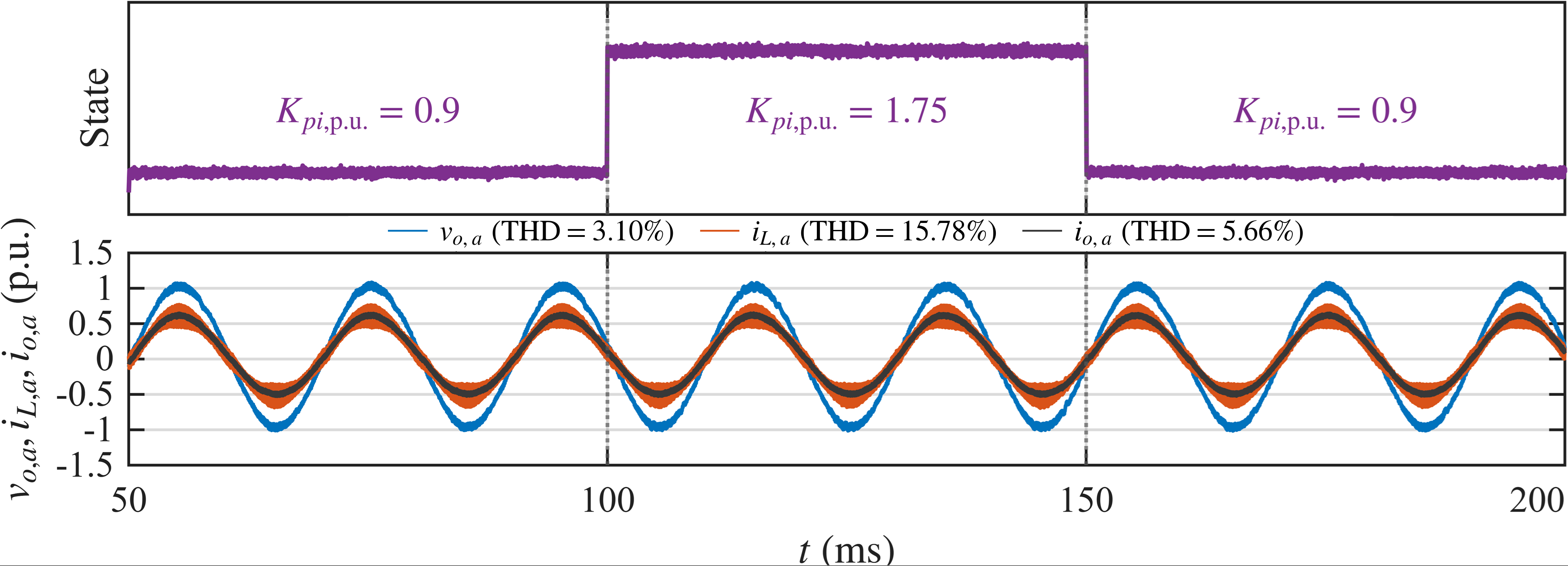}}
\par\vspace{1pt}
\subfloat[]{\includegraphics[width=3.20815in]{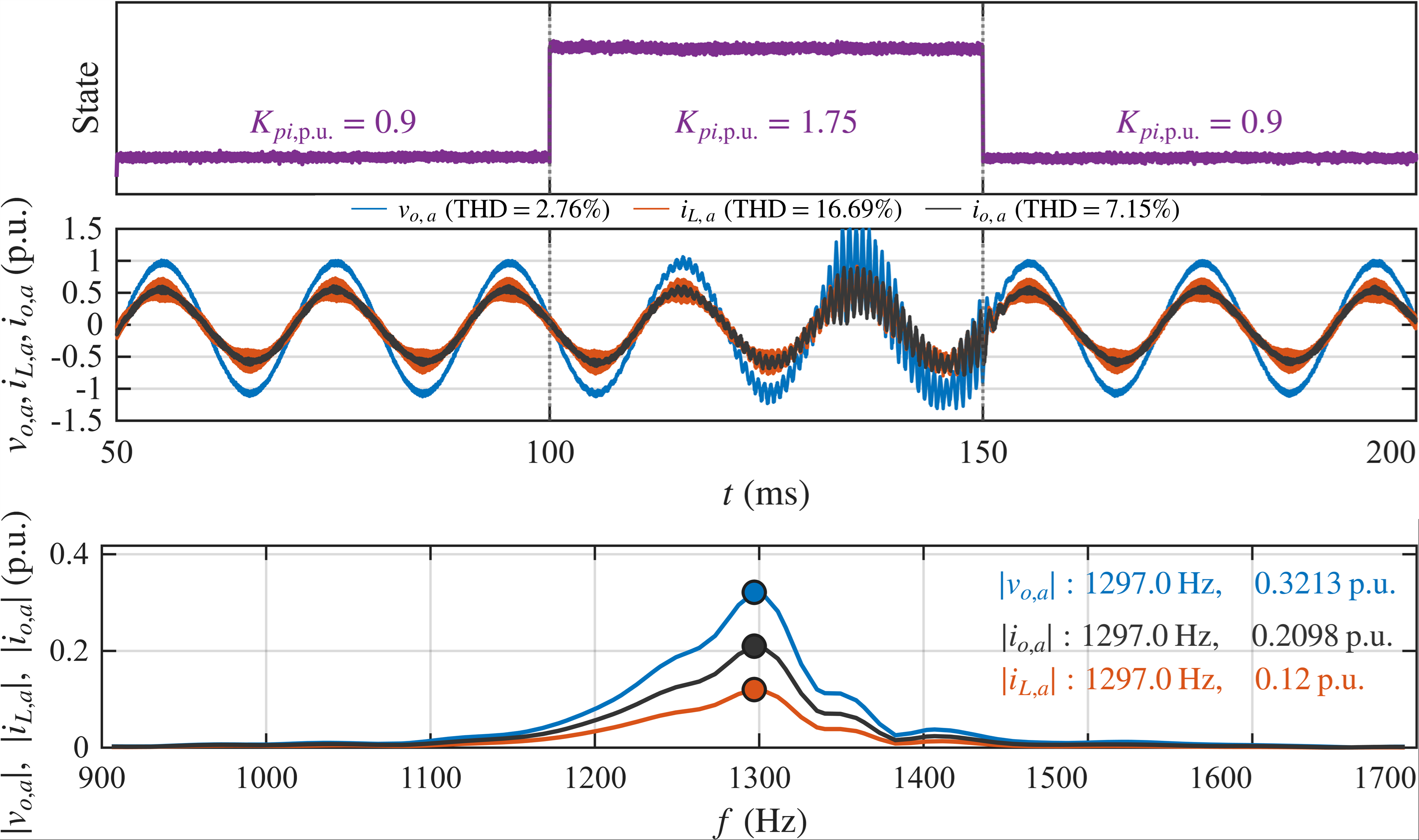}}
\caption{Grid-connected tests at 0.5 p.u. active power: (a) $\mathrm{SCR}=2.1$; (b) $\mathrm{SCR}=21$ and FFT results. THD: total harmonic distortion.}
\label{fig:17}
\end{figure}

Fig. 17 confirms stable operation at SCR = 2.1 and a 1.297-kHz oscillation at SCR = 21, 5.82\% above the continuous-model prediction. This corroborates the mechanism near the nominal boundary; implementation and parasitics can shift thresholds and modes. Restoring nominal gain suppresses the oscillation. Lower nominal grid-current distortion is consistent with capacitor diversion of high-frequency ripple. The nominal point illustrates operation near a grid boundary; practical gains require clearance within the common region.

\section{Conclusion}

This paper quantified how current gain constrains LC-resonance placement in unity-VFD, VA-based dual-loop grid-forming control. Increasing gain contracts the PVRR and can close the effective-gain window, which AD retuning alone cannot reopen. Simple proportional-path shaping recovers this window while retaining unity VFD. Reduced-model criteria, checked against the complete model, link window closure to the physical-resonance placement limit. Experiments demonstrate an 83.3\% LC-product reduction at comparable voltage and nominal current crossovers, and higher current gain with the same filter in a separate comparison. Grid-connected tests corroborate damping recovery and show that internal stability does not ensure stable grid interaction. The framework creates design freedom for coordinated filter selection and tuning, with clearance from internal and grid-interaction boundaries required for parameter and implementation variations.

\appendices
\setlength{\abovedisplayskip}{4pt plus 2pt minus 1pt}
\setlength{\belowdisplayskip}{4pt plus 2pt minus 1pt}
\setlength{\abovedisplayshortskip}{4pt plus 1pt minus 1pt}
\setlength{\belowdisplayshortskip}{4pt plus 1pt minus 1pt}
\setlength{\jot}{3pt}

\section{Current-Loop Models and Design Relations}

When the proportional path dominates, (11) gives the nominal current-loop crossover
\begin{equation}{\omega }_{c,i}\approx \frac{{\omega }_{\mathrm{LPF}}}{\sqrt{2}}{\left[ \sqrt{1+4{\left( \frac{{K}_{pi}}{{L}_{f}{\omega }_{\mathrm{LPF}}}\right) }^{2}}-1\right] }^{\frac{1}{2}}.\tag{A1}\label{eq:A1}\end{equation}
The corresponding nominal phase margin is
\begin{equation}{\mathrm{PM}}_{i}\approx{90}^{\circ}-\left[{\omega}_{c,i}{T}_{d}+\tan^{-1}\!({\omega}_{c,i}/{\omega}_{\mathrm{LPF}})\right]180^{\circ}/\pi.\tag{A2}\label{eq:A2}\end{equation}

\begin{samepage}
With VA feedback opened and VFD and AD retained, (12) gives
\begin{equation}T_{i,\mathrm{VFD\text{-}AD}}(s)=\frac{sC_fG_d(s)G_i(s)}{D_v(s)-sC_fG_d(s)G_i(s)}.\tag{A3}\label{eq:AiVFDAD}\end{equation}
\end{samepage}
Its closed-loop characteristic factor is $D_v(s)$, which forms the denominator of $T_v(s)$ in (12).

\section{Voltage-Loop Model and Output Impedance}

Substituting (2) into (4) and using $sL_fi_L=v_{\mathrm{inv}}-v_o$ gives
\begin{equation}\begin{aligned}
i_L(sL_f+G_dG_i)&={}G_dG_iY_{\mathrm{v}}e^*-\bigl[1-G_dH_v\\
&\quad+G_dG_i(Y_{\mathrm{v}}+K_vG_{\mathrm{N}})\bigr]v_o.
\end{aligned}\tag{B1}\label{eq:B1}\end{equation}
Under no-load conditions, $i_L=sC_fv_o$, yielding
\begin{equation}\begin{aligned}
\Delta(s)&={}sC_f(sL_f+G_dG_i)+1-G_dH_v\\
&\quad +G_dG_iK_vG_{\mathrm{N}}+G_dG_iY_{\mathrm{v}}.
\end{aligned}\tag{B2}\label{eq:B2}\end{equation}
With $D_v$ defined in (12), $\Delta=D_v+G_dG_iY_{\mathrm v}=D_v(1+T_v)$; closed-loop poles satisfy $\Delta=0$.

With $|G_d|=1$, applying $|T_v|=1$ to (12) gives
\begin{equation}|G_i(j\omega_{c,v})Y_{\mathrm v}(j\omega_{c,v})|=|D_v(j\omega_{c,v})|.\tag{B3}\label{eq:B3}\end{equation}
Select the voltage crossing below the nominal $f_{c,i}$, with $f_{c,v}=\omega_{c,v}/(2\pi)$. For $G_{\mathrm N}\approx1$ and dominant AD support, $D_v\approx G_dG_iK_v$. With inductive VA, $Y_{\mathrm v}\approx1/(sL_{\mathrm v})$, so $T_v\approx1/(sL_{\mathrm v}K_v)$ and $f_{c,v}\approx1/(2\pi L_{\mathrm v}K_v)$, recovering (10).
Using $\arg G_d=-\omega T_d$, the phase margin is
\begin{equation}\begin{aligned}
\mathrm{PM}_v&=180^\circ+\frac{180^\circ}{\pi}\bigl[\arg G_i(j\omega_{c,v})-\arg D_v(j\omega_{c,v})\\
&\quad-\omega_{c,v}T_d-\tan^{-1}(\omega_{c,v}L_{\mathrm v}/R_{\mathrm v})\bigr].
\end{aligned}\tag{B4}\label{eq:B4}\end{equation}
Phases use a consistent unwrapped branch.

Under the SISO assumptions of Section II-A, setting $e^*=0$ and substituting $i_o=i_L-sC_fv_o$ into (B1) gives (21), with $i_o$ positive toward the grid and $Z_o=-v_o/i_o$.

\section{Resonance-Band Crossing Relations}

Applying the resonance-band approximations in Section III-A and writing $G_d(j\omega)G_{\mathrm{LPF}}(j\omega)=Me^{-j\phi}$, (14) becomes
\begin{equation}\begin{aligned}
\Delta_v'(j\omega)&={}1-\omega^2L_fC_f-e^{-j\theta}\\
&\quad +Me^{-j\phi}\left[K_{\mathrm{eff}}+jK_{pi}\left(\omega C_f-\frac{1}{\omega L_{\mathrm{v}}}\right)\right].
\end{aligned}\tag{C1}\label{eq:C1}\end{equation}
With $B$ and $Q$ defined before (15), multiplying (C1) by $e^{j\phi}$ and separating real and imaginary parts gives
\begin{equation}\begin{aligned}
MK_{\mathrm{eff}}&=-(B_r\cos(\phi)-B_i\sin(\phi)),\\
MQ&=-(B_r\sin(\phi)+B_i\cos(\phi)).
\end{aligned}\tag{C2}\label{eq:C2}\end{equation}
Dividing (C2) by $M$ gives (15).

\microtypesetup{expansion=true,protrusion=false}
\raggedbottom
\renewcommand{\IEEEbibitemsep}{0pt}
\bibliographystyle{IEEEtran}
\bibliography{LC_Expan_references}

\makeatletter
\def\@IEEEBIOskipN{0.5\baselineskip}
\makeatother

\noindent\begin{minipage}{\columnwidth}
\begin{IEEEbiography}[{\includegraphics[width=1in,height=1.25in,keepaspectratio]{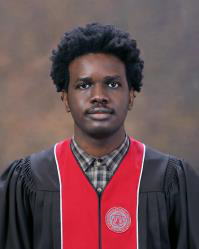}}]{Stephen Arinze Obi}
(Graduate Student Member, IEEE) received the B.S. and M.S. degrees from Kyungpook National University, Daegu, South Korea, in 2021 and 2023, respectively. He is currently pursuing the Ph.D. degree in electronic and electrical engineering at the same university. His research interests include power electronics and grid integration, renewable energy harvesting and storage, grid-following and grid-forming control, and control of power converters.
\end{IEEEbiography}

\begin{IEEEbiography}[{\includegraphics[width=1in,height=1.25in,keepaspectratio]{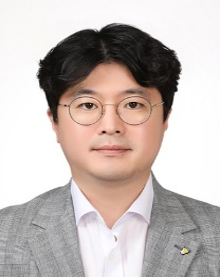}}]{Jae-Jung Jung}
(Senior Member, IEEE) received the B.S., M.S., and Ph.D. degrees in electrical engineering and computer science from Seoul National University, Seoul, South Korea, in 2011, 2013, and 2017, respectively. He was a Senior Engineer at Samsung Electronics, South Korea (2017--2019), and has been with Kyungpook National University, Daegu, South Korea, since 2019, where he is an Associate Professor in the Department of Electrical Engineering. He headed its Microgrid Research Center (2021--2025). Since 2022, he has chaired the Next-Generation eGrid Converter Research Council of the Korean Institute of Power Electronics and served as Treasurer of the IEEE Taegu Section PELS Chapter. His research covers grid-connected power electronics, electric-machine control, and power system dynamics.
\end{IEEEbiography}
\end{minipage}

\end{document}